\documentclass[twocolumn,tighten,times]{aastex63} 
\usepackage{lineno}

\usepackage{amsmath} 
\usepackage{amssymb}
\extrafloats{100}

\newcommand{\GP}{{\sc GALPROP}}

\newcommand{\galprop}{\textsc{GalProp}}

\newcommand{\fermilat}{\emph{Fermi}-LAT}

\newcommand{\hi}{H\,{\sc i}}
\newcommand{\hii}{H\,{\sc ii}}

\newcommand{\gray}{$\gamma$-ray}

\shorttitle{Geometric Origins of Radio/Infared/\gray{} Correlation}
\shortauthors{Porter et al.}

\begin{document}

\title{Geometry, Not Calorimetry, Drives the Radio/Infrared/\gray{} Correlation}

\author[0000-0002-2621-4440]{T.~A.~Porter} 
\affiliation{Hansen Experimental Physics Laboratory, Stanford University, Stanford, CA 94305}
\affiliation{Kavli Institute for Particle Astrophysics and Cosmology, Stanford University, Stanford, CA 94305}

\author[0000-0001-6141-458X]{I.~V.~Moskalenko} 
\affiliation{Hansen Experimental Physics Laboratory, Stanford University, Stanford, CA 94305}
\affiliation{Kavli Institute for Particle Astrophysics and Cosmology, Stanford University, Stanford, CA 94305}

\author[0000-0003-1458-7036]{G.~J\'{o}hannesson} 
\affiliation{Science Institute, University of Iceland, Dunhaga 3, IS-107 Reykjavik, Iceland}





\begin{abstract}
  We investigate whether the observed radio-infrared-\gray{} correlation in star-forming galaxies is a geometric effect rather than a signature of local cosmic-ray (CR) calorimetry.
  Using the \GP{} framework, we generate synthetic observations for external viewers from a grid of 3D Milky Way (MW) models with varied CR source, gas, interstellar radiation, and magnetic field distributions, all normalised to reproduce local CR data.
  We find that a tight, quasi-linear correlation arises naturally from line-of-sight integration through the extended, radially-structured disc, even when local calorimetry is absent.
  The correlation's properties depend strongly on viewing geometry, preserving its form under moderate inclination but breaking down in edge-on views where galactic components are stratified.
  We conclude that for a MW-like galaxy the correlation is primarily an emergent property of geometric projection and viewing angle, not a direct tracer of local calorimetry.
  While these findings directly apply to systems with similar structure and star formation properties, their extension to the diverse population of galaxies that form the global radio-infrared-\gray{} correlation requires further investigation.
  This geometric perspective implies that the scatter in these relations can be a powerful diagnostic of galactic structure and CR escape.
\end{abstract}


\keywords{Galactic cosmic rays (567); Secondary cosmic rays (1438)}

\section{Introduction} \label{Intro}

The tight, linear correlation between a galaxy's infrared (IR) and radio continuum (RC) luminosities is one of the most robust and ubiquitous scaling relations in extragalactic astronomy.
Established over three decades ago \citep{1985A&A...147L...6D, 1985ApJ...298L...7H}, this relation holds with remarkable consistency across several orders of magnitude in luminosity, spanning quiescent spirals, irregulars, and intense starburst galaxies \citep{1992ARA&A..30..575C, 2001ApJ...554..803Y}.
The canonical explanation invokes a shared energetic origin in massive star formation: ultraviolet radiation from young, hot stars is absorbed by interstellar dust and re-radiated in the IR, while the same stellar population ends in core-collapse supernovae that accelerate the cosmic-ray (CR) electrons responsible for synchrotron emission in the galaxy's magnetic fields.

The remarkable tightness of the RC-IR correlation, however, belies the complexity of the underlying physics.
The classic ``electron calorimeter'' model posits that in optically thick starburst environments, CR electrons lose all their energy via synchrotron and inverse Compton (IC) radiation before they can escape, naturally tying their total emission to the star formation rate (SFR) \citep{1989A&A...218...67V}.
Yet, the correlation persists even in low-density, CR-transparent galaxies where electrons are expected to escape, suggesting that a simple calorimetric argument is insufficient.
This has led to the proposal of scenarios where various factors, including CR injection and transport, magnetic field amplification, and gas density, are coupled in a self-regulating manner to maintain the correlation across diverse galactic environments \citep[e.g.,][]{2010ApJ...717....1L,2021MNRAS.502.1312C}.

High-energy \gray{} observations provide an additional and independent probe of these processes. 
The expectation of detectable \gray{} flux from extragalactic star-forming regions \citep[e.g.,][]{1989A&A...213L..12V} was first confirmed by EGRET's detection of the Large Magellanic Cloud (LMC) \citep{1992ApJ...400L..67S}.
However, the instrument's limited sensitivity yielded only upper limits for more distant starbursts like NGC 253 and M82, as well as M31 \citep{1999ApJ...516..744B}.
The greatly enhanced sensitivity of the \textit{Fermi} Large Area Telescope (\textit{Fermi}-LAT) expanded the sample of detected star-forming galaxies sufficiently to permit statistical studies for the first time.
Analysis of \textit{Fermi}-LAT data for a sample of 64 galaxies, selected on the basis of their molecular gas content along with the \gray{} detected Local Group members, demonstrated a quasi-linear scaling between the \gray{} luminosity and the SFR \citep{2012ApJ...755..164A}.
The extension to include \gray{s} is significant because it traces both CR electrons and, uniquely, the dominant CR nuclei population via hadronic interactions. The radio-IR-\gray{} correlation therefore links the total CR energy budget to the multi-wavelength properties of the host galaxy.

The study of nearby galaxies, whose large angular sizes permit resolved multi-wavelength analysis, allows the focus to shift from galaxy-integrated properties to sub-galactic scales.
In these systems, such as the M31, M33, and the Magellanic Clouds, the RC-IR correlation has been shown to persist down to scales of $\sim$100~pc, albeit with increased scatter and local variations in the correlation slope \citep{2006ApJ...638..157M, 2006MNRAS.370..363H, 2013A&A...557A.129T}.

The inclusion of \gray{s}, however, introduces a fundamental scale ambiguity.
While a few key systems,  the LMC, SMC, and M31, are spatially resolved in \gray{s} the angular resolution is significantly poorer than in the radio and IR bands, probing physical scales closer to a kiloparsec than 100 pc.
Beyond these few cases, the broader sample of star-forming galaxies remains unresolved in \gray{s}.
This leaves a critical gap in our understanding: the well-established local ($\sim$100 pc) RC-IR correlation exists in a regime where the \gray{} emission is either poorly resolved or completely unresolved.
It is therefore unclear whether the local RC-IR correlation reflects a direct physical coupling (i.e., calorimetry) that would extend to \gray{s} if we had sufficient resolution, or if it is simply a higher-resolution manifestation of the same geometric projection effects that govern the global scaling relations.
The distinction is critical, as the answer determines whether these emissions can be used as reliable, local tracers of star formation or are instead smoothed by CR propagation over kiloparsec scales.

In this paper, we argue that geometry, not local calorimetry, is the principal driver of the apparent universality of the radio-IR-\gray{} correlation on sub-galactic scales.
We hypothesise that the observed tight scaling is largely a consequence of line-of-sight (LOS) integration through extended, but fundamentally correlated, large-scale distributions of CRs and their interstellar medium (ISM) target densities (gas, radiation, and magnetic fields).
To test this, we employ the \galprop\ CR propagation framework \citep{2022ApJS..262...30P} to construct a suite of three-dimensional (3D), physically-motivated models of the Milky Way (MW).
By systematically varying the spatial distributions of CR sources, interstellar gas, radiation fields, and magnetic field configurations, we self-consistently compute the resulting synchrotron, IR, and \gray{} emission.
We then generate synthetic, resolved maps as seen from a range of observer locations, distances, and viewing inclinations.
This approach allows us to disentangle the effects of intrinsic physical coupling from those of observational geometry, revealing how the act of observation itself can forge tight correlations from spatially complex and imperfectly coupled components.
Our work underscores the need to account for geometric projection effects when interpreting these multi-wavelength correlations, with direct implications for their use as star formation tracers and as probes of CR physics in the MW and beyond.

\section{\GP\ Framework}
The \GP{} framework is a widely used, publicly available numerical tool for modeling Galactic CR transport and the associated diffuse non-thermal emission.
The central idea in \galprop{} is that many different observables trace the same underlying CR populations and Galactic environment, and should therefore be modelled together.
These include direct measurements of CR nuclei, $\bar{p}$, and $e^\pm$, as well as diffuse \gray{} emission and synchrotron radiation. 
Its range of physical validity extends from sub-keV\,--\,PeV energies for particles and from $10^{-6}$ eV ($\mu$eV)\,--\,PeV for photons.
The aim is to build models that are as realistic as the data allow, using available astrophysical, nuclear, and particle-physics inputs and making key modelling assumptions explicit so they can be tested via controlled variations \citep{2007ARNPS..57..285S}.
  
\GP{} has 27+ years of development behind it \citep{1998ApJ...493..694M, 1998ApJ...509..212S}.
It is consistently updated to keep up with the ever increasing amount and precision of experimental data.
The latest release of \galprop{}, version 57.1, became public in 2022 \citep[][and references therein]{2022ApJS..262...30P}.
The dedicated website\footnote{https://galprop.stanford.edu \label{site}} provides source releases, supporting data products and run configurations for reproducing published results, and a facility to run \GP\ releases via web browsers \citep{2011CoPhC.182.1156V}.

The \GP{} code solves a system of about 90+ time-dependent transport equations (partial differential equations in 3D or 4D: spatial variables plus energy) with a given source distribution and boundary conditions to give the intensity distributions for all CR species through the ISM: $^1$H$-$\,$^{64}$Ni, $\bar{p}$, $e^\pm$. 
The spatial boundary conditions assume free particle escape. 
The propagation equations include terms for convection, distributed reacceleration, energy losses, nuclear fragmentation, radioactive decay, and production of secondary particles and isotopes \citep[for details of these processes and formalism, see][]{1998ApJ...509..212S}. 
 
For a given halo size, the diffusion coefficient $D_{xx}(\rho)$, as a function of rigidity $\rho$, and other propagation parameters can be determined from secondary-to-primary nuclei ratios, typically B/C, [Sc+Ti+V]/Fe, and/or $\bar{p}/p$. 
If reacceleration is included, the momentum-space diffusion coefficient $D_{pp}$ is related to the spatial coefficient $D_{xx} = \beta D_0 \rho^\delta$ \citep{1994ApJ...431..705S}, where $\beta=v/c$ is the particle velocity, $\rho$ is the magnetic rigidity, and $\delta = 1/3$ for a Kolmogorov spectrum of interstellar turbulence \citep{1941DoSSR..30..301K}, or $\delta = 1/2$ for an Iroshnikov$-$Kraichnan cascade \citep{1964SvA.....7..566I, 1965PhFl....8.1385K}, but can also be arbitrary.
The spatial diffusion coefficient can also depend on position. 
This option was developed and utilised by \citet{2015ApJ...799...86A}, where the positional dependence of the spatial diffusion coefficient was linked to the distribution of the GMF strength.

CR source distributions can be specified using a composition scheme that allows the spatial density distribution, spectral characteristics, and respective contributions to be customised.
Possible components for the spatial density model include an axisymmetric disc, spiral arms, various central bulges, and other structures.
  Each basic component can be further split up and fine-tuned with different radial profiles, so that different classes of sources have their own population spatial distribution, injection spectra, and isotopic abundances, allowing for a very flexible description of a galaxy.

The injection spectra of CR species for a source density distribution are parameterised by a multiple broken power law in rigidity:
\begin{equation}
q(\rho) \propto (\rho/\rho_0)^{-\gamma_0}\prod_{i=0}^N\left[1 + (\rho/\rho_i)^\frac{\gamma_i - \gamma_{i+1}}{s_i}\right]^{s_i},
\label{eq:injection}
\end{equation}
where $\gamma_{i =0,\dots,N+1}$ are the spectral indices, $\rho_{i = 0,\dots,N}$ are the break rigidities, and $s_i$ are the smoothing parameters ($s_i$ is negative/positive for $|\gamma_i |\lessgtr |\gamma_{i+1} |$). Each primary isotope can have unique spectral parameters. 

The \GP{} code computes a complete network of primary, secondary, and tertiary isotope production starting from input CR source abundances.
The nuclear reaction network is built using the \citet{2018NDS...151D...3.}, where a detailed description of its method of construction is given by \citet{2020ApJS..250...27B,BoDe21}. 
Included are multistage chains of $p$, $n$, $d$, $t$, $^3$He, $\alpha$, and $\beta^\pm$-decays, and K-electron capture, as well as, in several cases, more complicated reactions.
This accounts for up to 81 daughter nuclei in the final state for each fragment produced in the spallation of the target nucleus, plus an unlimited number of $p$, $n$, and $\beta^\pm$-decays. 

Because the decay branching ratios and half-lives of fully stripped and hydrogen-like ions may differ (a well-known example is $^7$Be), \galprop{} includes the processes of K-electron capture, electron pickup from the neutral ISM gas, and formation of hydrogen-like ions as well as the inverse process of electron stripping \citep{1973RvMP...45..273P, 1978PhDT........12W, 1979PhDT........67C}. Meanwhile, the fully stripped and hydrogen-like ions are treated as separate species.

The isotopic production cross section routines are built using LANL nuclear codes \citep{2001ICRC....5.1836M, 2003ICRC....4.1969M, 2004AdSpR..34.1288M, 2005AIPC..769.1612M, 1996PhRvC..54.1341B, 1998PhRvC..57..233B}, databases (LANL, EXFOR\footnote{https://www-nds.iaea.org/exfor/}), an extensive literature search \citep{BoDe21}, and parameterizations \citep{2003ApJS..144..153W, 1998ApJ...501..911S, 1998ApJ...501..920T, 1983ApJS...51..271L}. 
Cross sections for production of $^2$H, $^3$H, $^3$He, and total inelastic cross sections parameterisations \citep{1996PhRvC..54.1329W, BarPol1994} and corrected Tripathi formalism are described in \citet[][Appendices F, G]{2022ApJS..262...30P}. 
The latest updates of the production cross sections are summarised in \citet[][Appendix B.4]{Cummings_2025}.

For the CR interactions with the interstellar gas, \galprop{} runs can use different density models. 
The ISM gas consists mostly of H and He with a ratio of 10:1 by number \citep{2001RvMP...73.1031F}.
Hydrogen can be found in the different states, atomic (\hi), molecular (H$_2$), or ionised (\hii), while He is mostly neutral.
\hi{} is $\sim$60\% of the mass, while H$_2$ and \hii{} contain 25\% and 15\%, respectively \citep{2001RvMP...73.1031F}.
The \hii{} gas has a low number density and scale height $\sim$few~100~pc.
The H$_2$ gas is clumpy and forms high-density molecular clouds.

For 2D calculations, we adopt the analytical model for the gas density distribution implemented in \GP{} (GP2D) \citep{2002ApJ...565..280M}.
The radial distribution for \hi{} is taken from \citet{1976ApJ...208..346G} while the vertical distribution is from \citet{1990ARA&A..28..215D} for galactocentric radial distances $0\le R\le 8$~kpc and \citet{1986A&A...155..380C} for $R\ge10$~kpc with linear interpolation in between.
The CO gas distribution is taken from \citet{1988ApJ...324..248B} for 1.5~kpc$<$$R$$<$10~kpc, and from \citet{1990A&A...230...21W} for $R$$\ge$10~kpc, and is augmented with the \cite{2007A&A...467..611F} model for $R$$\le$1.5 kpc.
Both of the 2D atomic and molecular gas distributions are rescaled to the common IAU recommended Sun-Galactic centre (GC) distance of $R_\odot = 8.5$~kpc \citep{1986MNRAS.221.1023K}.
The \hii{} gas distribution is given by the NE2001 model \citep{2002astro.ph..7156C, 2003astro.ph..1598C, 2004ASPC..317..211C} with the updates from \citet{2008PASA...25..184G}.

For 3D simulations, the \hi\ and $^{12}$CO distributions from \cite{2018ApJ...856...45J} (J18) are available.
These were developed using a maximum-likelihood forward-folding optimisation applied to the LAB-\hi\ \citep{2005A&A...440..775K} and CfA composite CO data \citep{2001ApJ...547..792D, 2004ASPC..317...66D}.
A recent development \citep{2026ApJ..1001...17P} is that the gas distribution from \citet{2025A&A...693A.139S} (S25) is now also available.
The S25 distribution is a reconstruction of a very similar collection of data to the J18 work using 3D Gaussian processes to model correlations in the interstellar gas over different sight lines to enforce coherent structure for its distribution.
Compared to the 2D models, the added degrees of freedom allow the optimised distributions to better reproduce the features observed in the line-emission surveys.

The 3D interstellar radiation field (ISRF) models for the MW were developed by \citet{2017ApJ...846...67P} based on spatially smooth stellar and dust models.
The ISRF models have designations R12 and F98 that correspond to the respective references supplying the stellar/dust distributions \citep{2012A&A...545A..39R,1998ApJ...492..495F}.
Both the R12 and F98 models provide equivalent solutions for the ISRF intensity distribution throughout the MW, but neither gives an overall best match with the data.
Toward the inner Galaxy, where the ISRF intensity is most uncertain, they are giving lower and upper bounds for its distribution, as determined by pair-absorption effects on sources toward the GC \citep[][]{2018PhRvD..98d1302P}.

The Galactic magnetic field (GMF) consists of the large-scale regular \citep{1985A&A...153...17B} and small-scale random \citep[e.g.,][]{2008A&A...477..573S} components that are about equal in intensity.
The random fields are mostly produced by the supernovae and other outflows,  which result in randomly oriented fields with a typical spatial scale of $\lesssim$100~pc \citep{1995MNRAS.277.1243G, 2008ApJ...680..362H}.
Also, there may be the anisotropic random (``striated'') fields, which refer to a large-scale ordering originating from stretching or compression of the random field \citep{2001SSRv...99..243B}.
This component is expected to be aligned to the large-scale regular field, with frequent reversal of its direction on small scales.
\galprop{} includes multiple large-scale MW GMF models \citep{2008A&A...477..573S, 2010MNRAS.401.1013J, 2010RAA....10.1287S, 2011ApJ...738..192P, 2012ApJ...757...14J}. 

\section{Modelling Setup}

To disentangle the effects of CR transport physics from geometric projection, we constructed a grid of physically-motivated MW models within the \GP{} framework. The grid is designed to systematically probe the impact of four key Galactic components: the CR source distribution, the interstellar gas, the ISRF, and the GMF.

The spatial dependence of the CR injection is explored using three source distributions that span a plausible range of scenarios.
Our baseline is a smooth, axisymmetric model derived from the observed pulsar distribution \citep[SA0;][]{2004A&A...422..545Y}.
We contrast this with models where sources are increasingly tied to the Galaxy's spiral structure, using hybrid (SA50) and pure-spiral (SA100) distributions from \citet{2017ApJ...846...67P}.
For the ISM target distribution, we likewise select representative models to bracket current uncertainties.
The gas is modelled either as the 2D axisymmetric GP2D distribution or the more realistic 3D J18 reconstruction.
For the ISRF, we use the F98 and R12 models.
The GMF is represented by either a simple 2D model with an exponential profile and a randomised field orientation ($B_\mathcal{E}$), or a more complex 3D model incorporating the coherent field from \citet{2011ApJ...738..192P} plus a random component ($B_\mathcal{P}$).

A critical aspect of our methodology is the normalisation procedure, which ensures that any differences in predicted emission are due to the large-scale component distributions.
For this, we followed the two-part optimisation procedure detailed in \citet{2022ApJS..262...30P}.
We first establish a single benchmark model by performing a global optimisation against a comprehensive set of local CR nuclei data (Be, B, C, O, etc.).
Then, for every other configuration in our grid, we re-tune the propagation parameters and injection spectra with the specific goal of reproducing the unmodulated local CR spectra of that benchmark model to within 5\%.
This isolates the effects of large-scale structure by ensuring all models are indistinguishable to an observer at the Solar position.

All calculations were performed on a 3D spatial grid, centred on the Galactic Centre (GC) with the Sun at $(X,Y,Z)=(8.5, 0, 0)$~kpc.
We employed the tangent grid function described in \citet{2022ApJS..262...30P}, which provides enhanced resolution of $\sim$50~pc in the solar neighbourhood and $\sim$25~pc vertically in the plane, coarsening toward the model boundaries at a $X_{\rm max}/Y_{\rm max}$ of $\pm$20~kpc and a halo height of $|Z_{\rm halo}| = 6$~kpc. The kinetic energy grid spans 10~MeV to 1~PeV over 32 logarithmic steps.

With the CR populations determined for each normalised model, the broadband non-thermal emissivities were calculated by folding the CR solutions with their respective ISM target distributions.
Hadronic \gray{} emission from protons and heavier nuclei, leptonic emission from IC and bremsstrahlung, and synchrotron radiation were all computed self-consistently, including the contribution from secondary electrons and positrons produced in hadronic interactions.
Secondary production from hadronic processes is treated using the cross sections from AAfrag202, a new option first used examining self-consistent diffuse neutrino emissions by \citet{2026ApJ..1004...30M}, which will be available in the next release of the \GP\ framework. 

\section{Results}

\begin{figure*}[t!]
  \centering
  \includegraphics[height=0.47\textwidth]{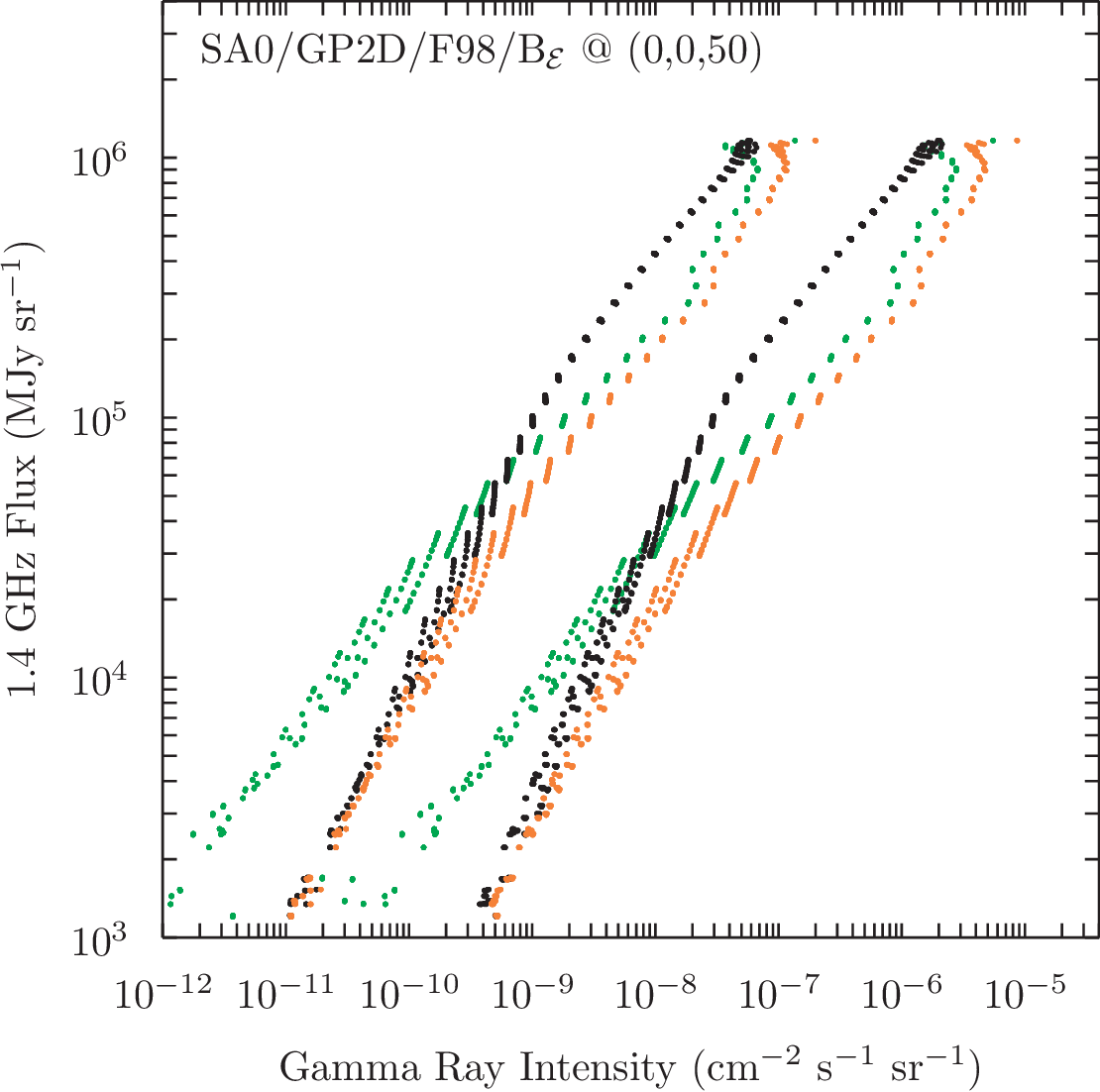}\hfill 
  \includegraphics[height=0.47\textwidth]{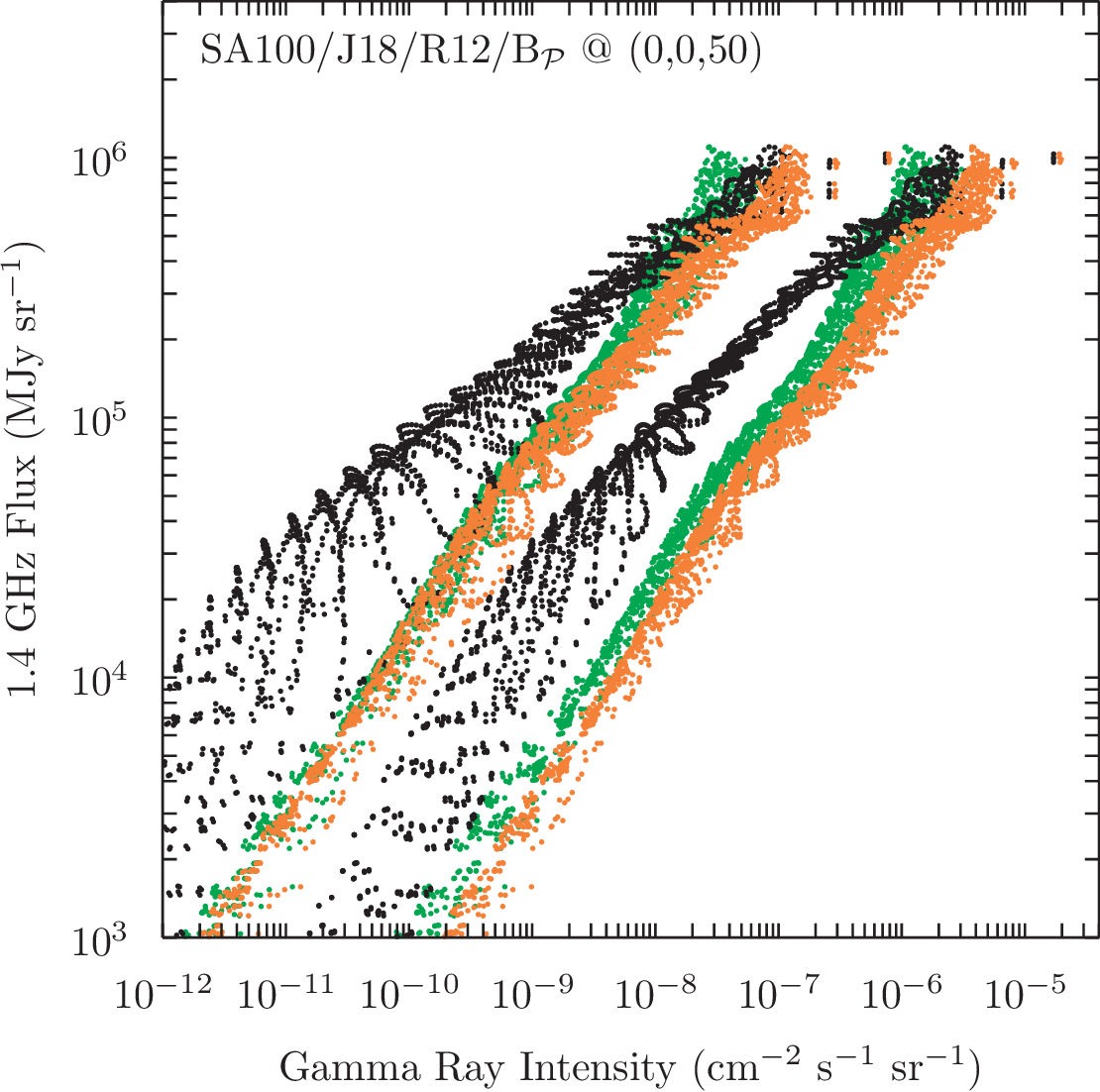}\hfill  \\ 
  \includegraphics[height=0.47\textwidth]{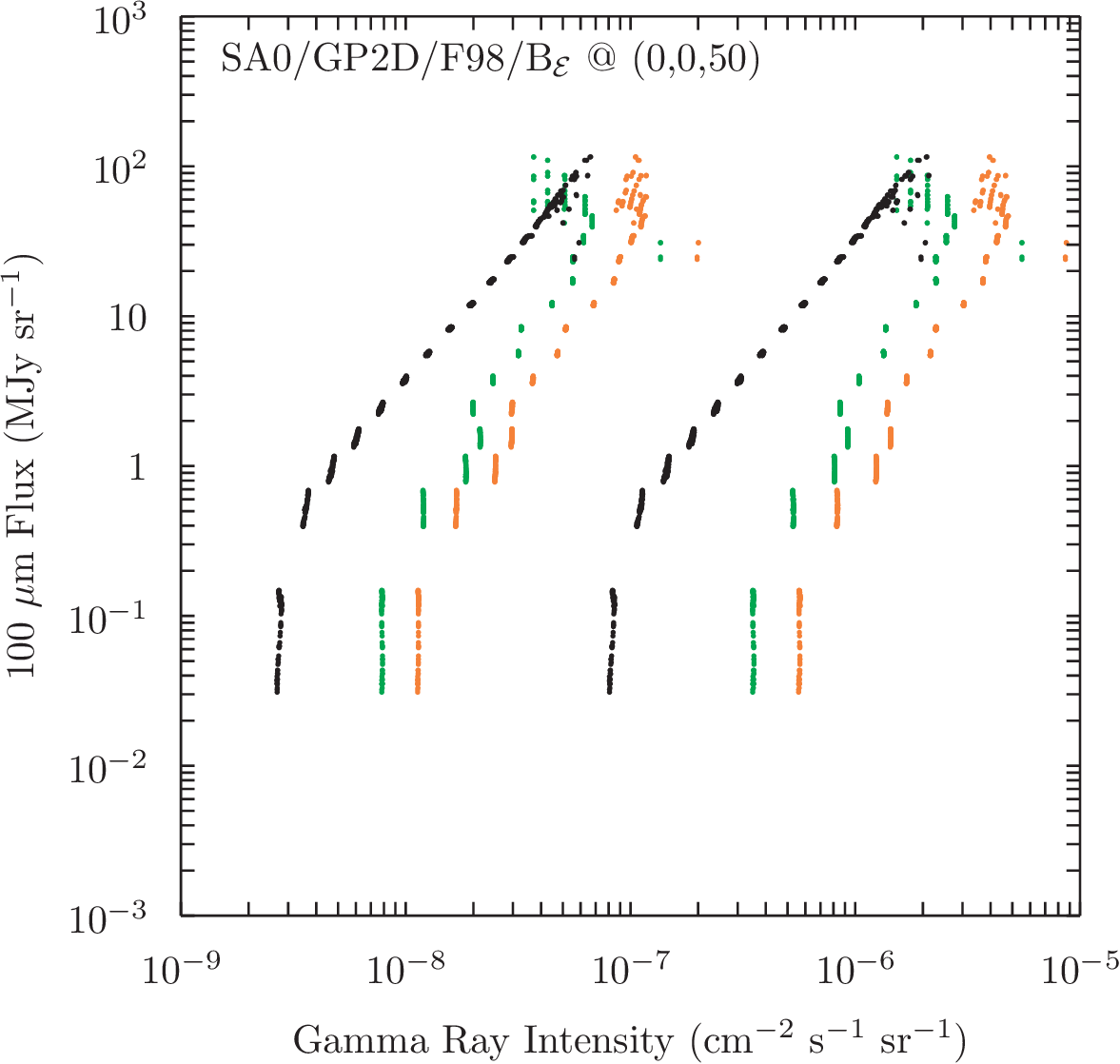}\hfill 
  \includegraphics[height=0.47\textwidth]{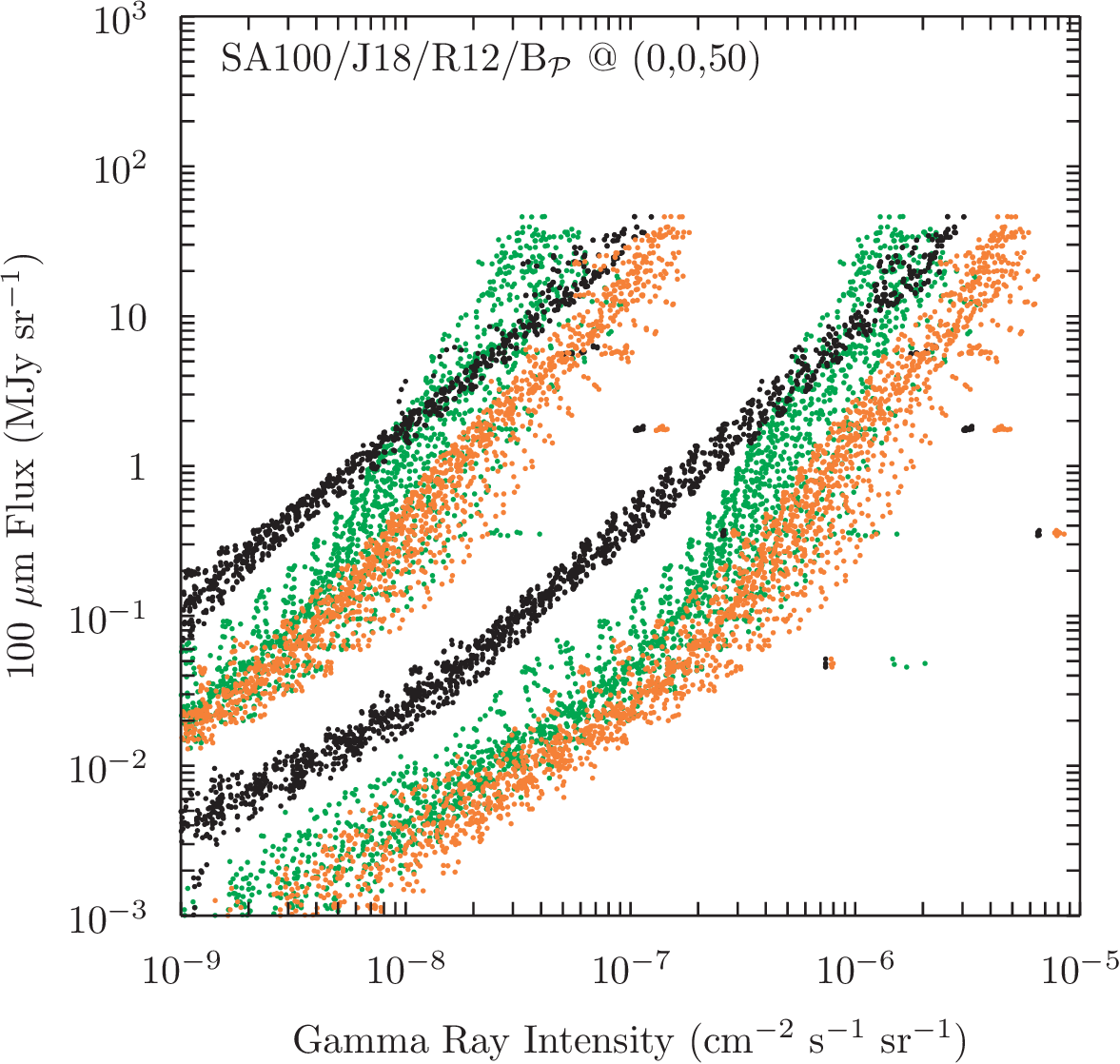}\hfill
  \caption{Correlation plots of (top) 1.4~GHz RC and (bottom) 100$\mu$m IR vs.\ \gray{} intensity for the SA0-GP2D-F98-B$_\mathcal{E}$ (left) and SA100-J18-R12-B$_\mathcal{P}$ (right) modelling configurations for an observer located at 50~kpc looking facedown toward the GC of the MW.
    For the \gray{} intensities, the correlations are separated into production processes: gas/$\pi^0$-decay, green; ISRF/Compton, black; total, orange. For each of these, the left point cloud corresponds to 10--100~GeV \gray{s} and the right to 1--10~GeV \gray{s}. \label{fig:ext_obs50kpc_hp6}}
\end{figure*}

Figure~\ref{fig:ext_obs50kpc_hp6} shows the multi-wavelength correlations for two contrasting model configurations: the near-axisymmetric case (SA0-GP2D-F98-B$_\mathcal{E}$, left column), which we will take as our benchmark configuration, and the most highly structured case (SA100-J18-R12-B$_\mathcal{P}$, right column) with CR sources  and other ISM components with non-axisymmetric  features, such as spiral arms, etc.
These synthetic observations are made from a face-on perspective at a distance of 50~kpc directly above the GC.
We chose this observer location because it is the approximate distance for the best-resolved external galaxy (LMC) for the RC, IR, and \gray{} bands.

For this figure, the LOS integrated intensity maps were made using a two-step process on a HEALPix grid.
The emission was first calculated at resolution $N_{\rm side}$=512, corresponding to a spatial scale of $\sim$100~pc at the observer's location.
This initial resolution was chosen to match the intrinsic resolving power of our input components, namely the ISRF models and the J18 gas reconstruction.
Subsequently, these maps were averaged to $N_{\rm side}$=128 to approximate the angular resolution of an instrument like the \fermilat\ at energies $\gtrsim$1~GeV, which represents the least-resolved band in our multi-wavelength comparison.

\begin{figure*}[t!]
  \centering
  \includegraphics[height=0.47\textwidth]{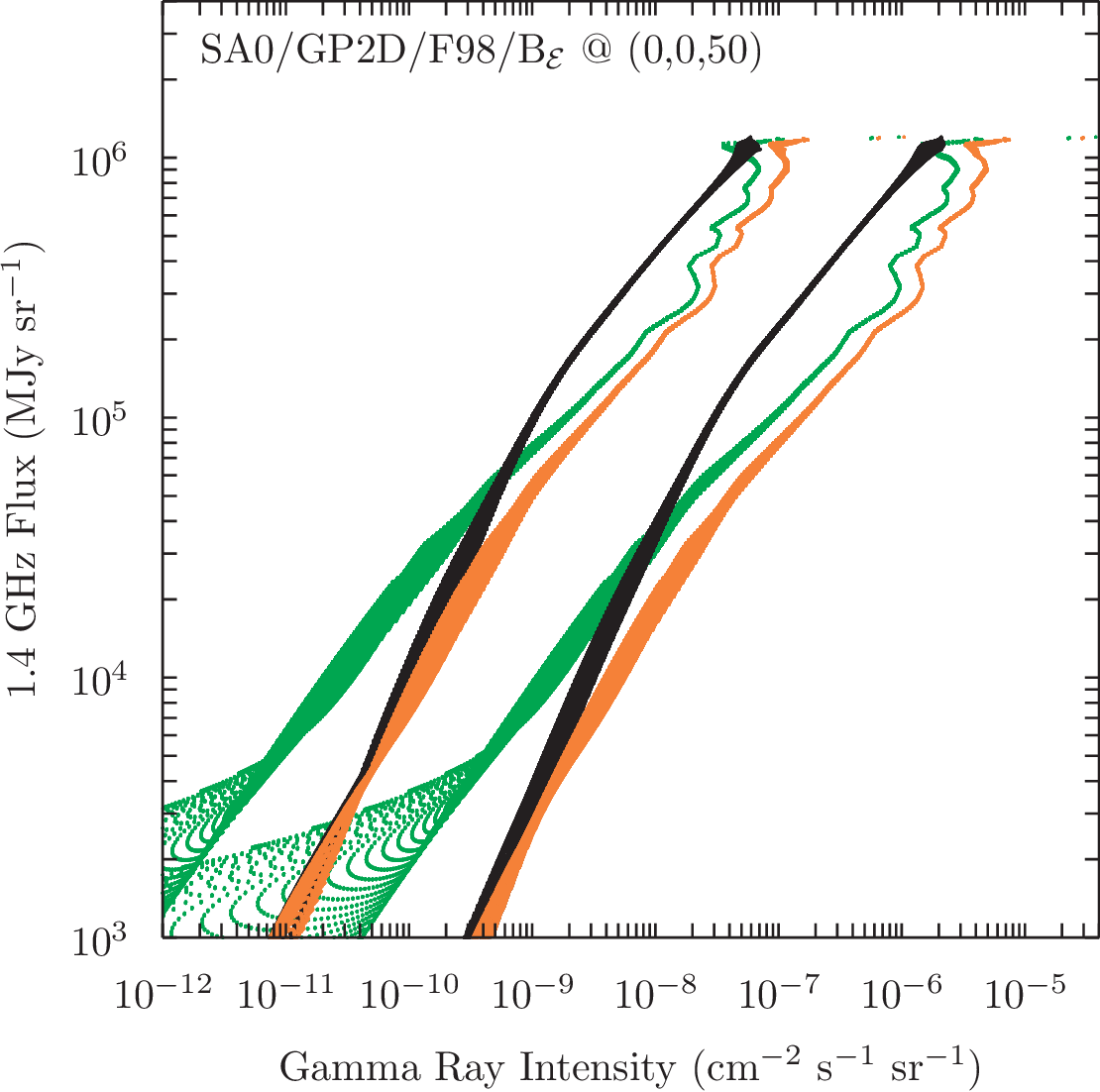}\hfill 
  \includegraphics[height=0.47\textwidth]{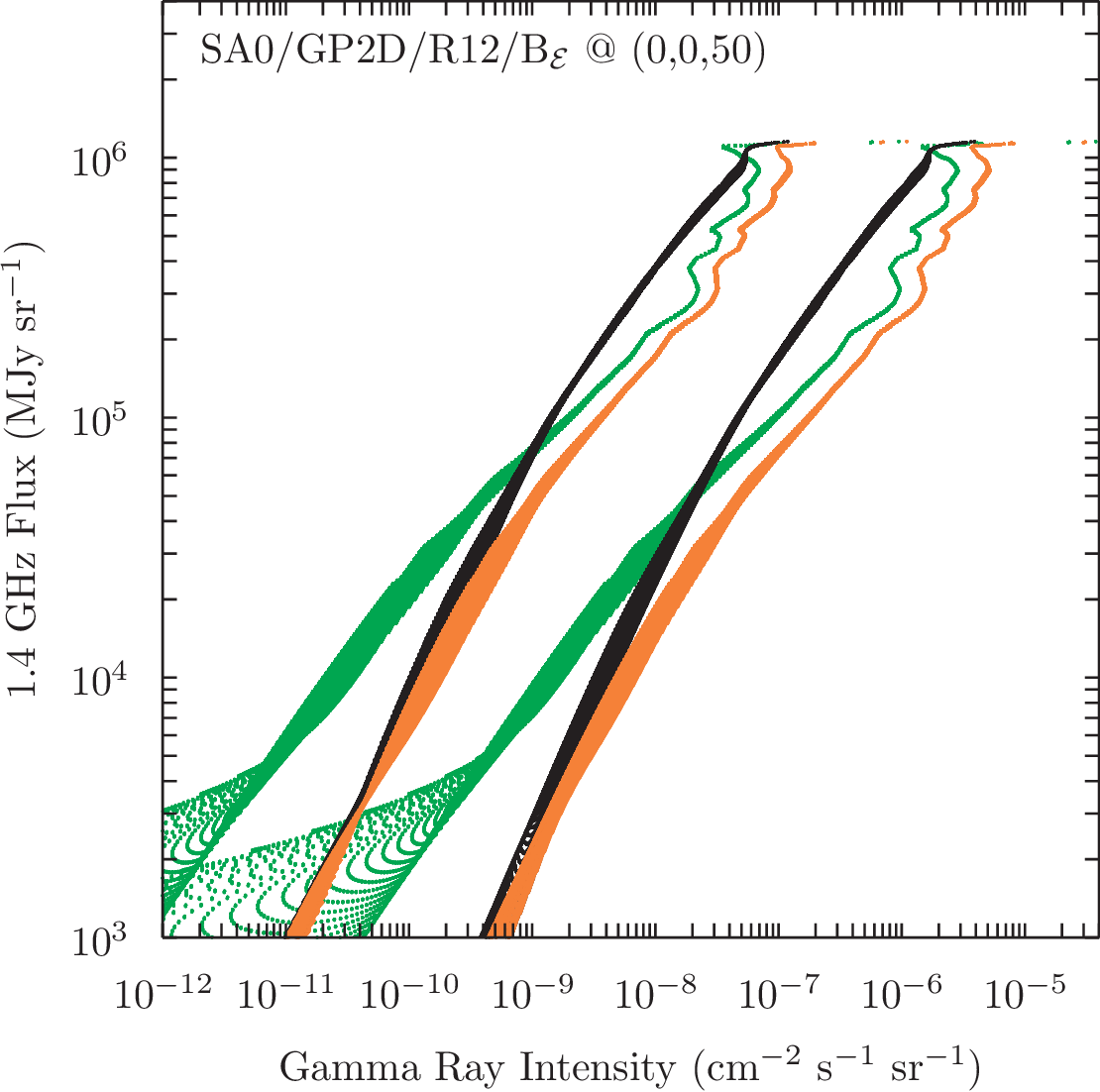}\hfill  \\ 
  \includegraphics[height=0.47\textwidth]{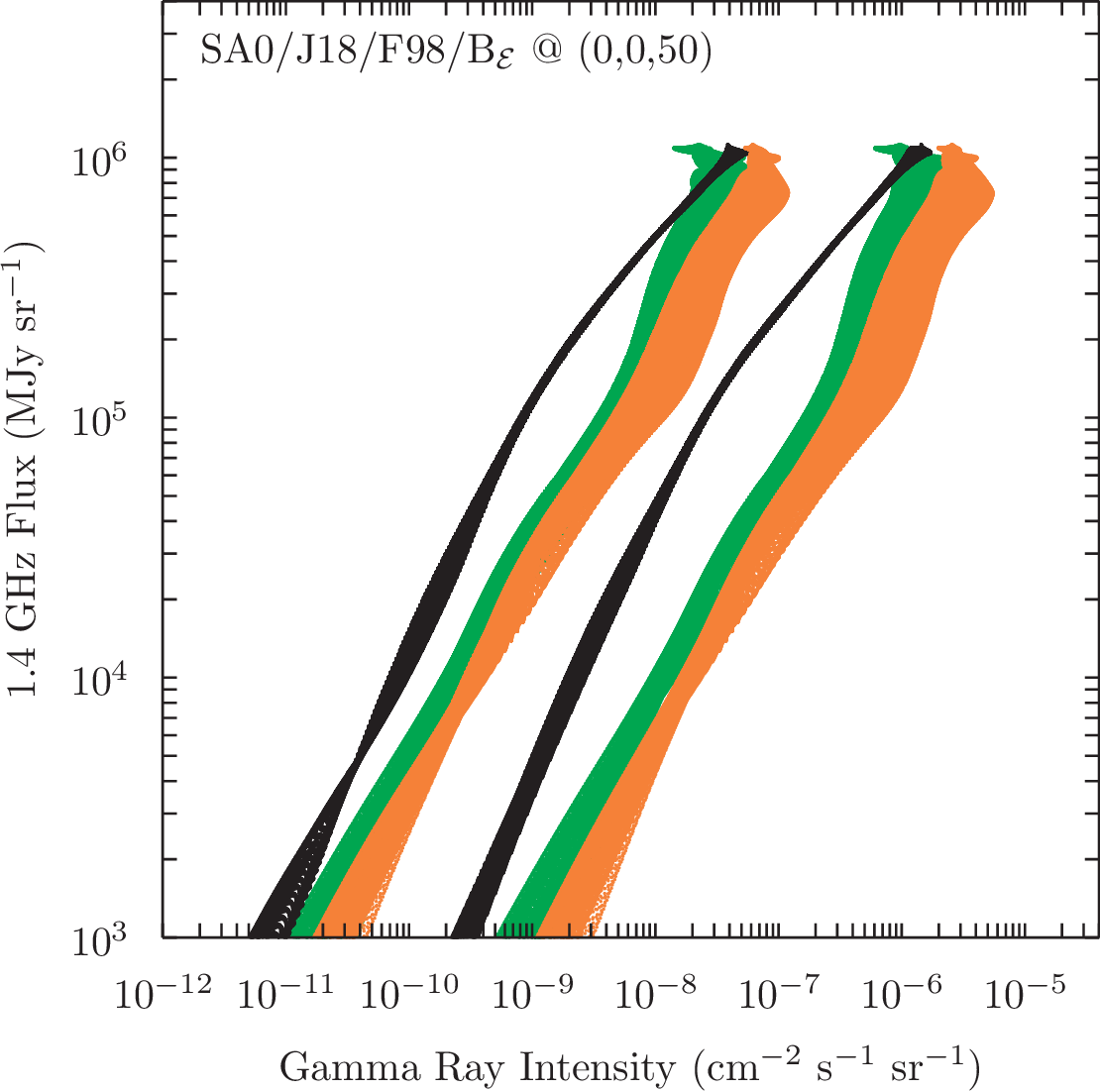}\hfill 
  \includegraphics[height=0.47\textwidth]{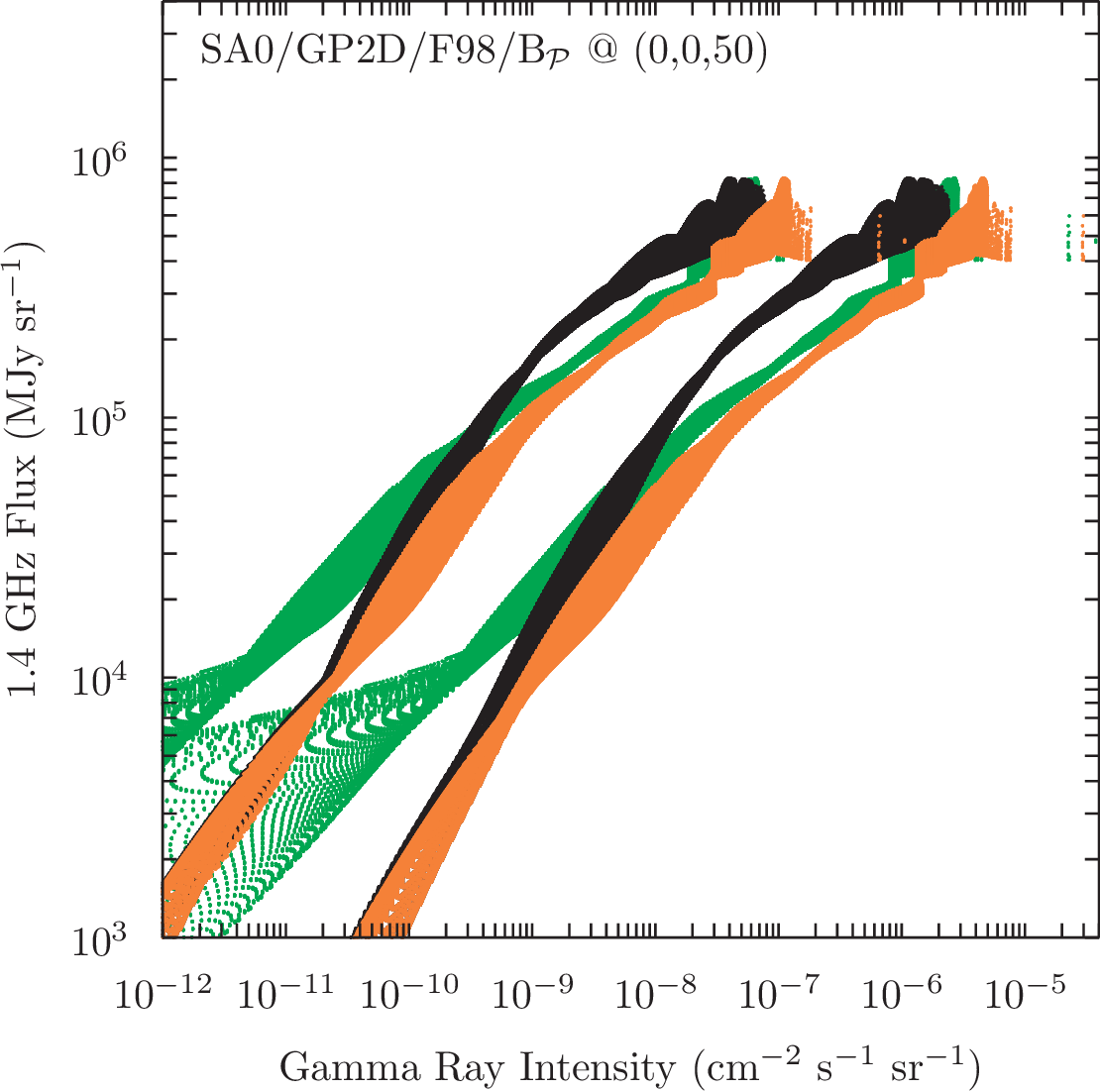}\hfill
  \caption{Correlation plots of (top) 1.4~GHz RC vs.\ \gray{} intensity for the 
benchmark SA0-GP2D-F98-B$_\mathcal{E}$ configuration (top left) with variations in the individual Galactic components: F98$\to$R12 (top right), GP2D$\to$J18 (bottom left), and B$_\mathcal{E}$$\to$B$_\mathcal{P}$ (bottom right)
for an observer located at 50~kpc looking facedown toward the GC of the MW.
For the \gray{} intensities, the correlations are separated into production processes: gas/$\pi^0$-decay, green; ISRF/Compton, black; total, orange. For each of these, the left point cloud corresponds to 10--100~GeV \gray{s} and the right to 1--10~GeV \gray{s}.  \label{fig:ext_obs50kpc_hp9_radiogamma}}
\end{figure*}

\begin{figure*}[t!]
  \centering
  \includegraphics[height=0.47\textwidth]{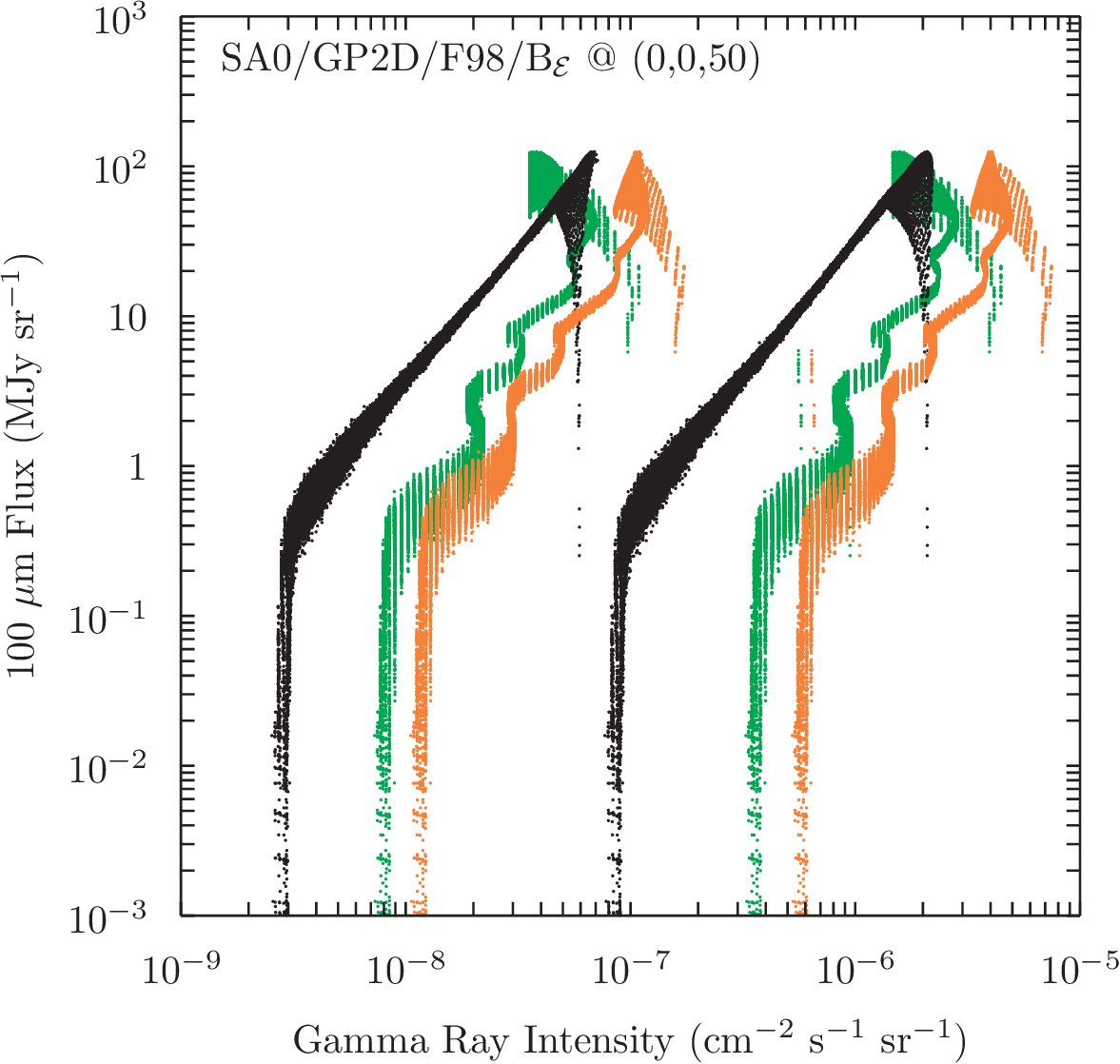}\hfill 
  \includegraphics[height=0.47\textwidth]{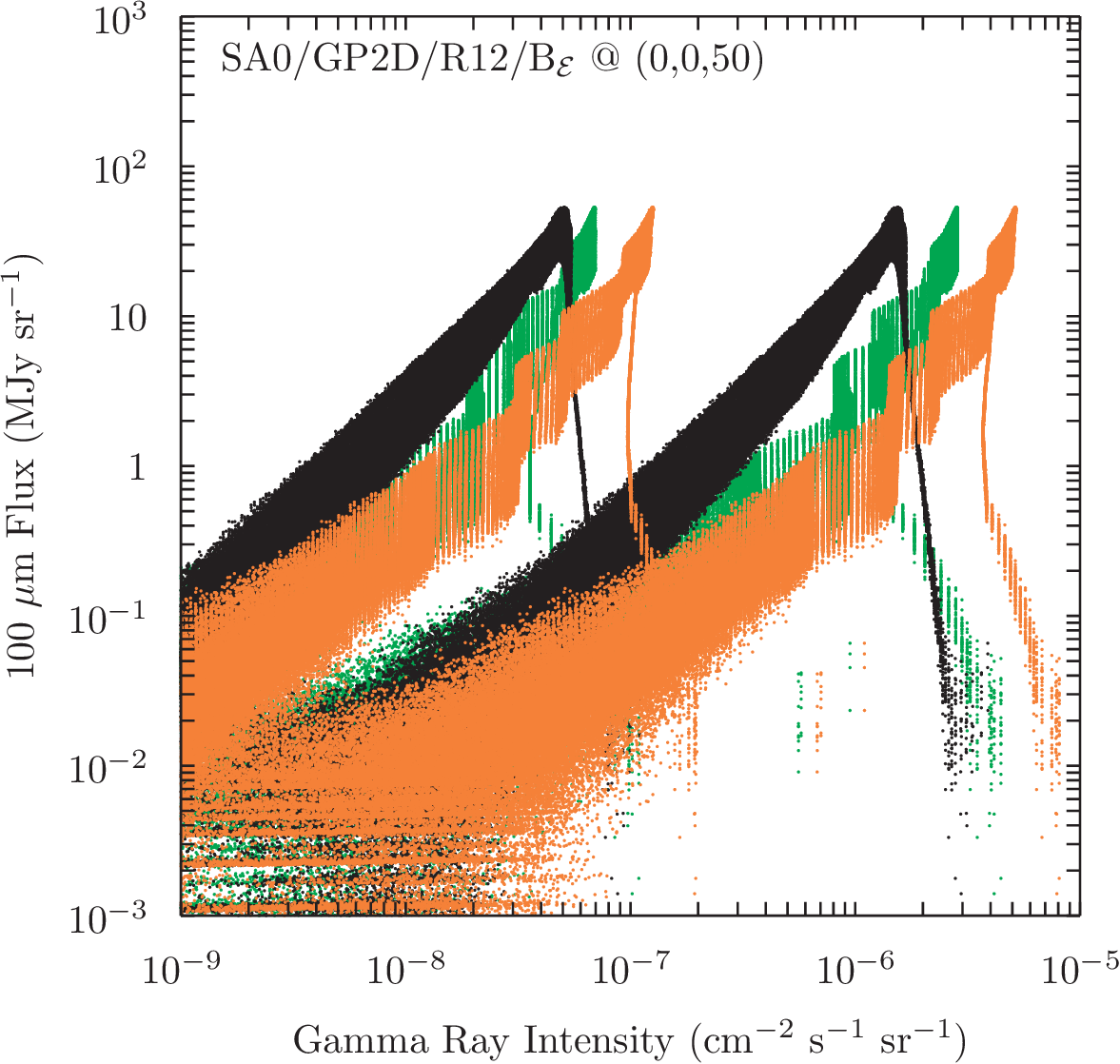}\hfill  \\ 
  \includegraphics[height=0.47\textwidth]{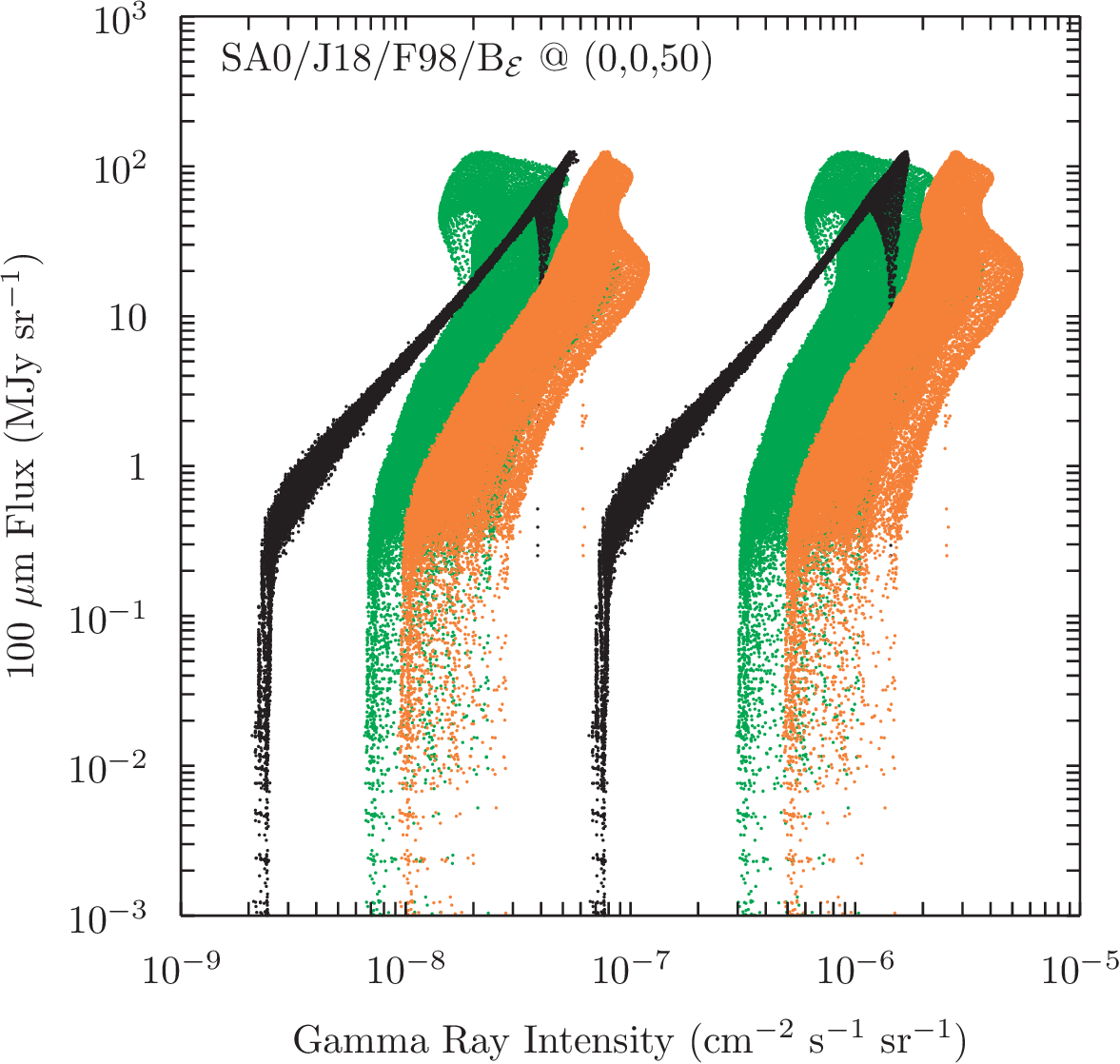}\hfill 
  \includegraphics[height=0.47\textwidth]{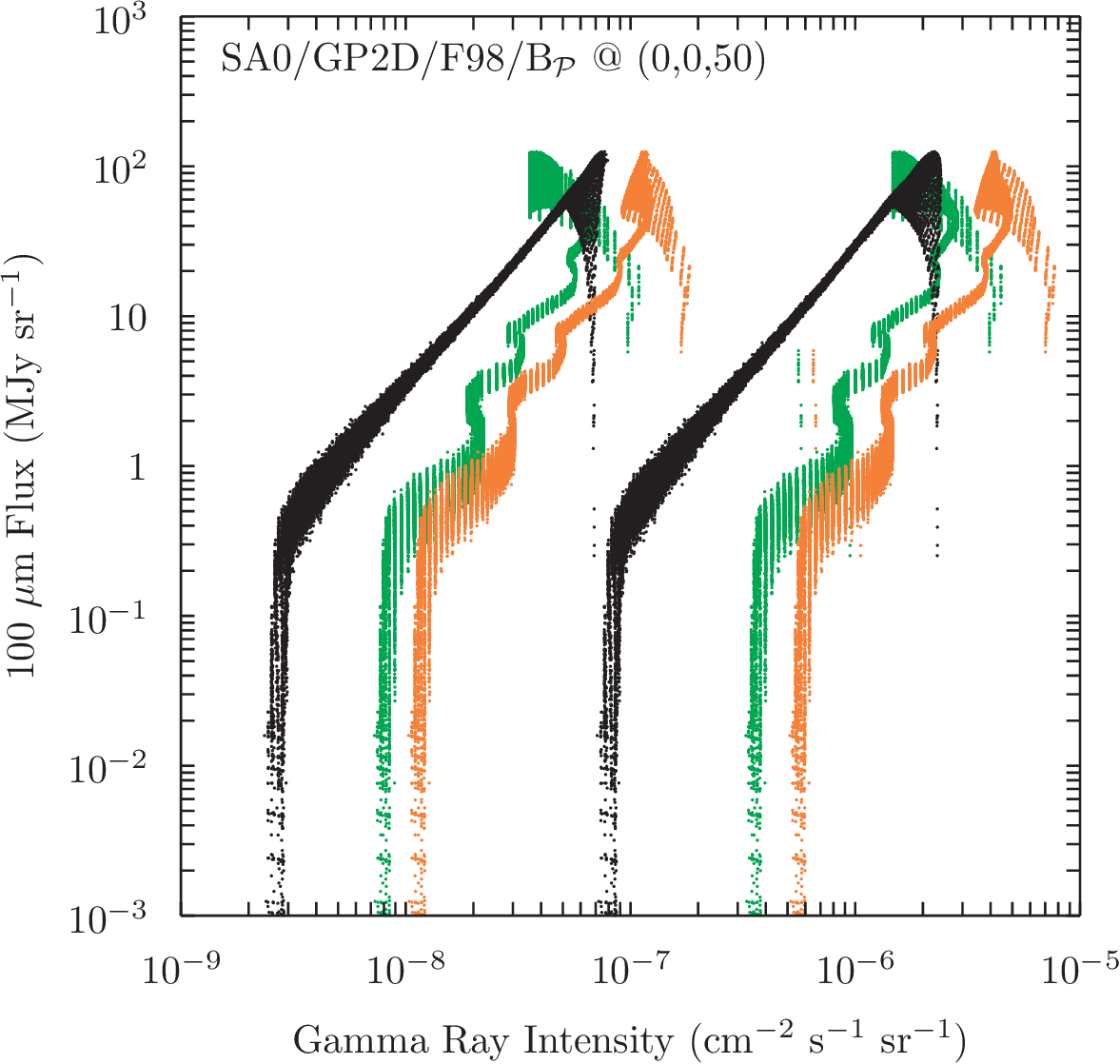}\hfill
  \caption{
Correlation plots of (top) 100~$\mu$m IR flux vs.\ \gray{} intensity for the 
benchmark SA0-GP2D-F98-B$_\mathcal{E}$ configuration (top left) with variations in the individual Galactic components: F98$\to$R12 (top right), GP2D$\to$J18 (bottom left), and B$_\mathcal{E}$$\to$B$_\mathcal{P}$ (bottom right)
for an observer located at 50~kpc looking facedown toward the GC of the MW.
For the \gray{} intensities, the correlations are separated into production processes: gas/$\pi^0$-decay, green; ISRF/Compton, black; total, orange. For each of these, the left point cloud corresponds to 10--100~GeV \gray{s} and the right to 1--10~GeV \gray{s}. \label{fig:ext_obs50kpc_hp9_irgamma}}
\end{figure*}

\begin{figure*}[t!]
  \centering
  \includegraphics[height=0.47\textwidth]{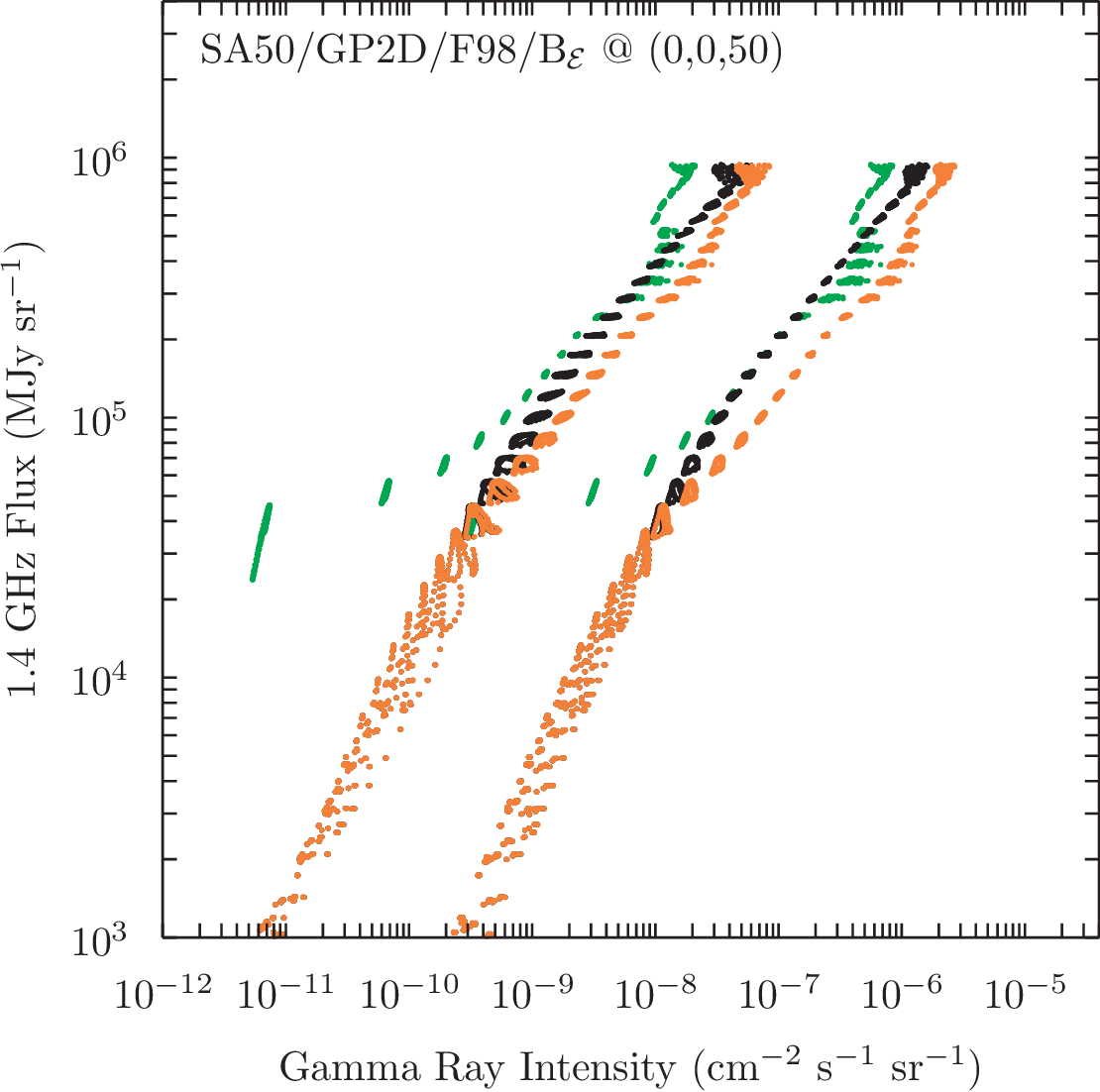}\hfill 
  \includegraphics[height=0.47\textwidth]{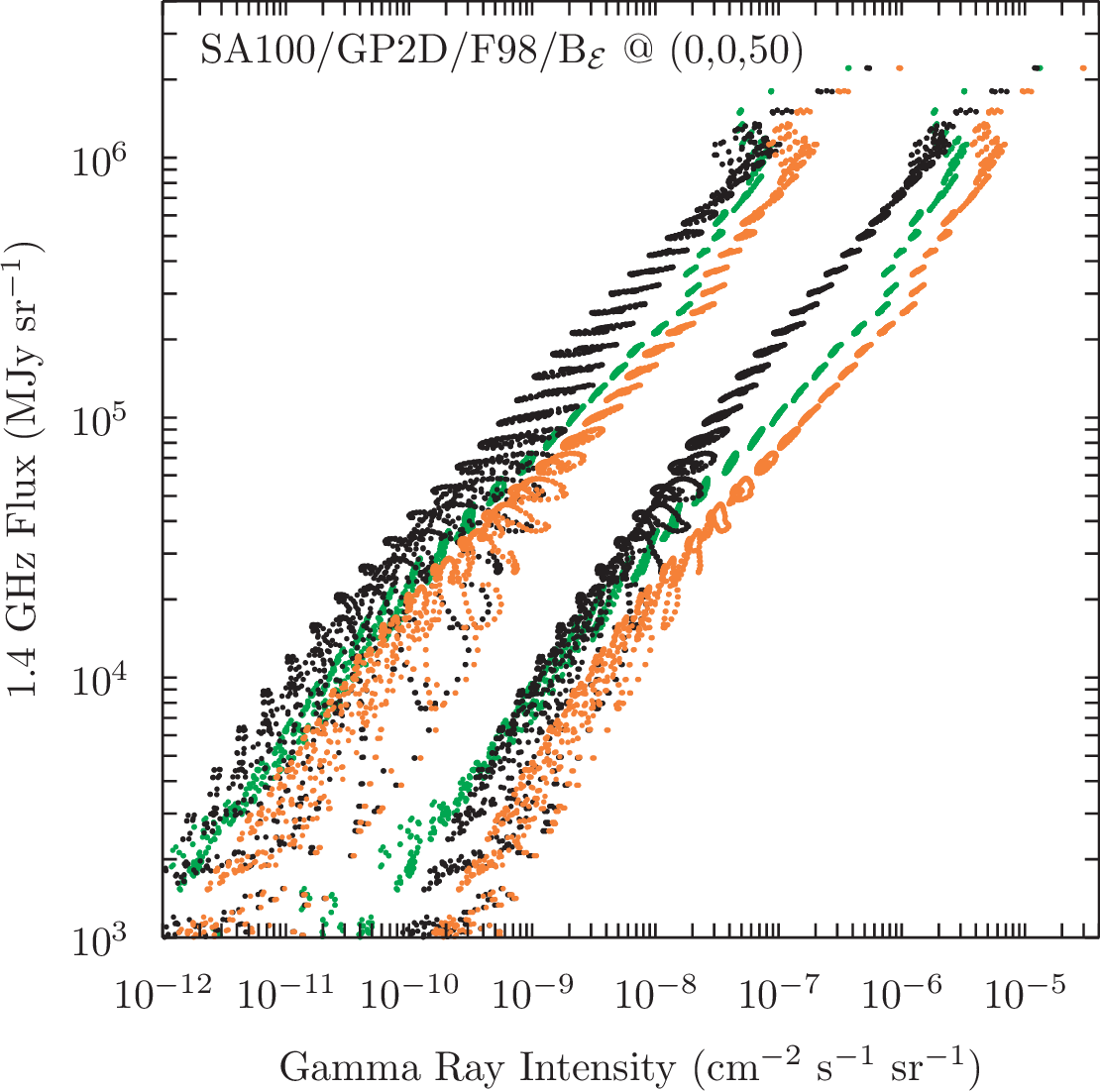}\hfill \\
  \includegraphics[height=0.47\textwidth]{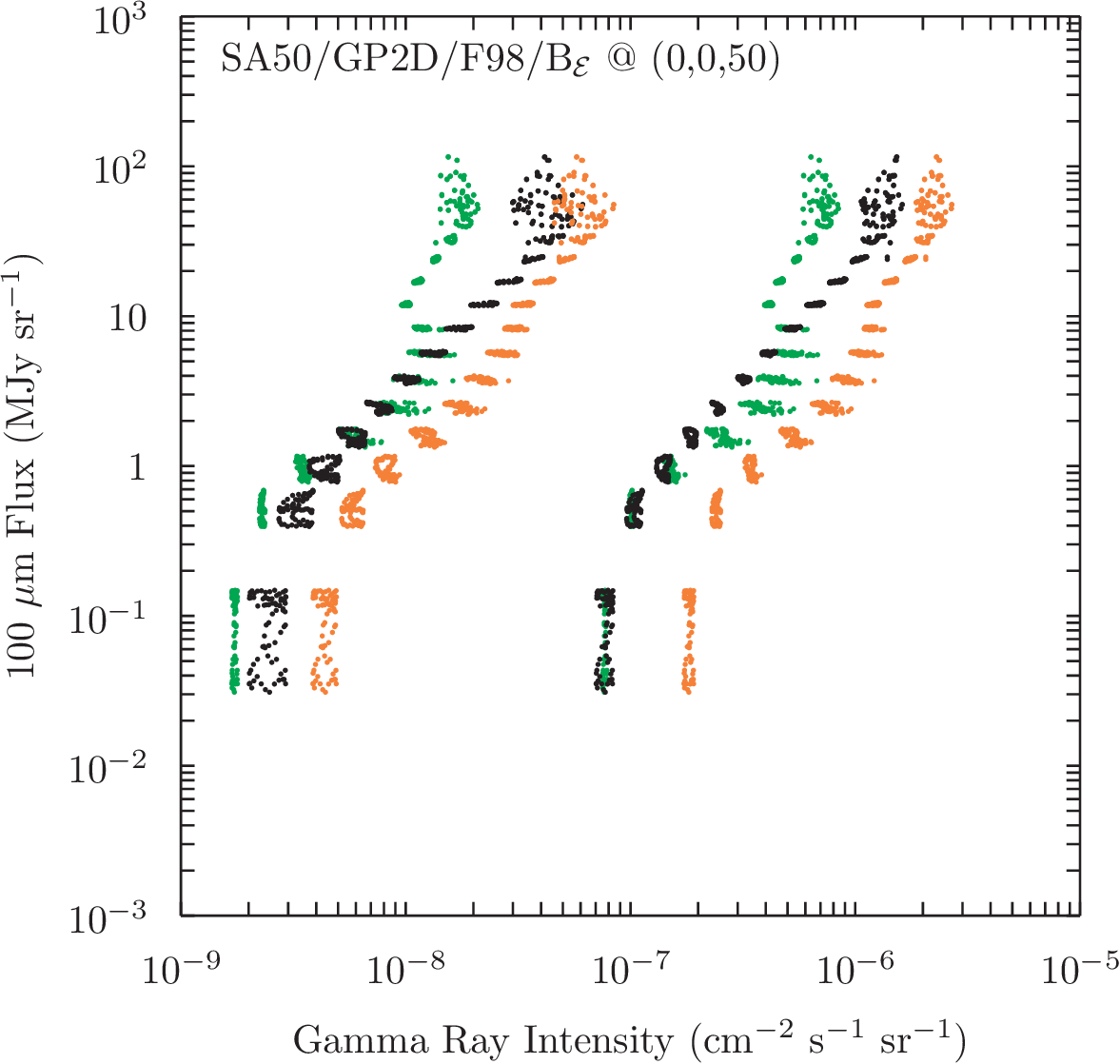}\hfill 
  \includegraphics[height=0.47\textwidth]{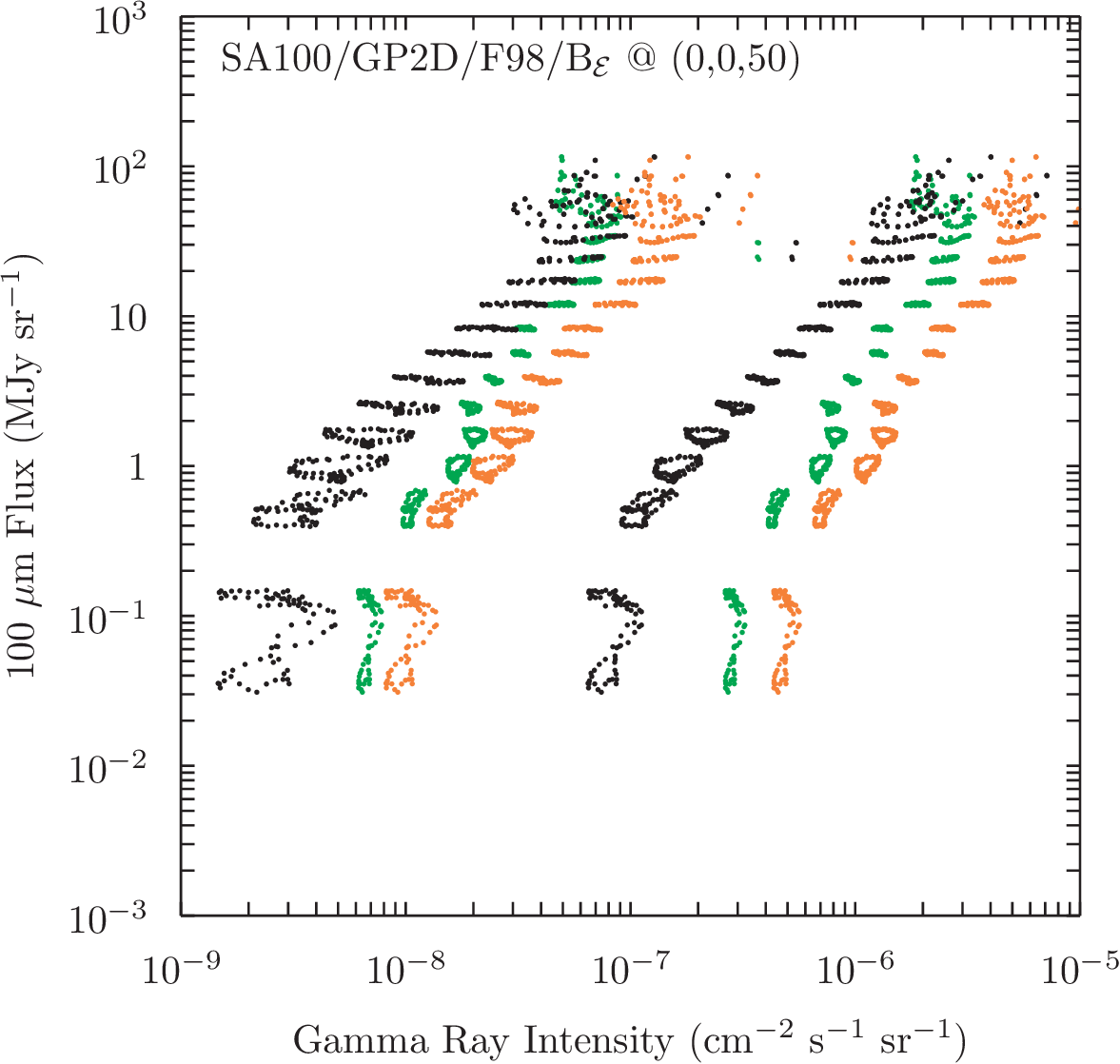}\hfill   
  \caption{Correlation plots of (top) 1.4~GHz RC and (bottom) 100~$\mu$m IR flux vs. \gray{} intensity for the SA50-GP2D-F98-B$_\mathcal{E}$ (left) and SA100-GP2D-F98-B$_\mathcal{E}$ (right) modelling configurations for an observer located at 50~kpc looking facedown toward the GC of the MW.
For the \gray{} intensities, the correlations are separated into production processes: gas/$\pi^0$-decay, green; ISRF/Compton, black; total, orange. For each of these, the left point cloud corresponds to 10--100~GeV \gray{s} and the right to 1--10~GeV \gray{s}.
    \label{fig:ext_obs50kpc_hp6_srcvar}}
\end{figure*}

\begin{figure*}[t!]
  \centering
  \includegraphics[height=0.47\textwidth]{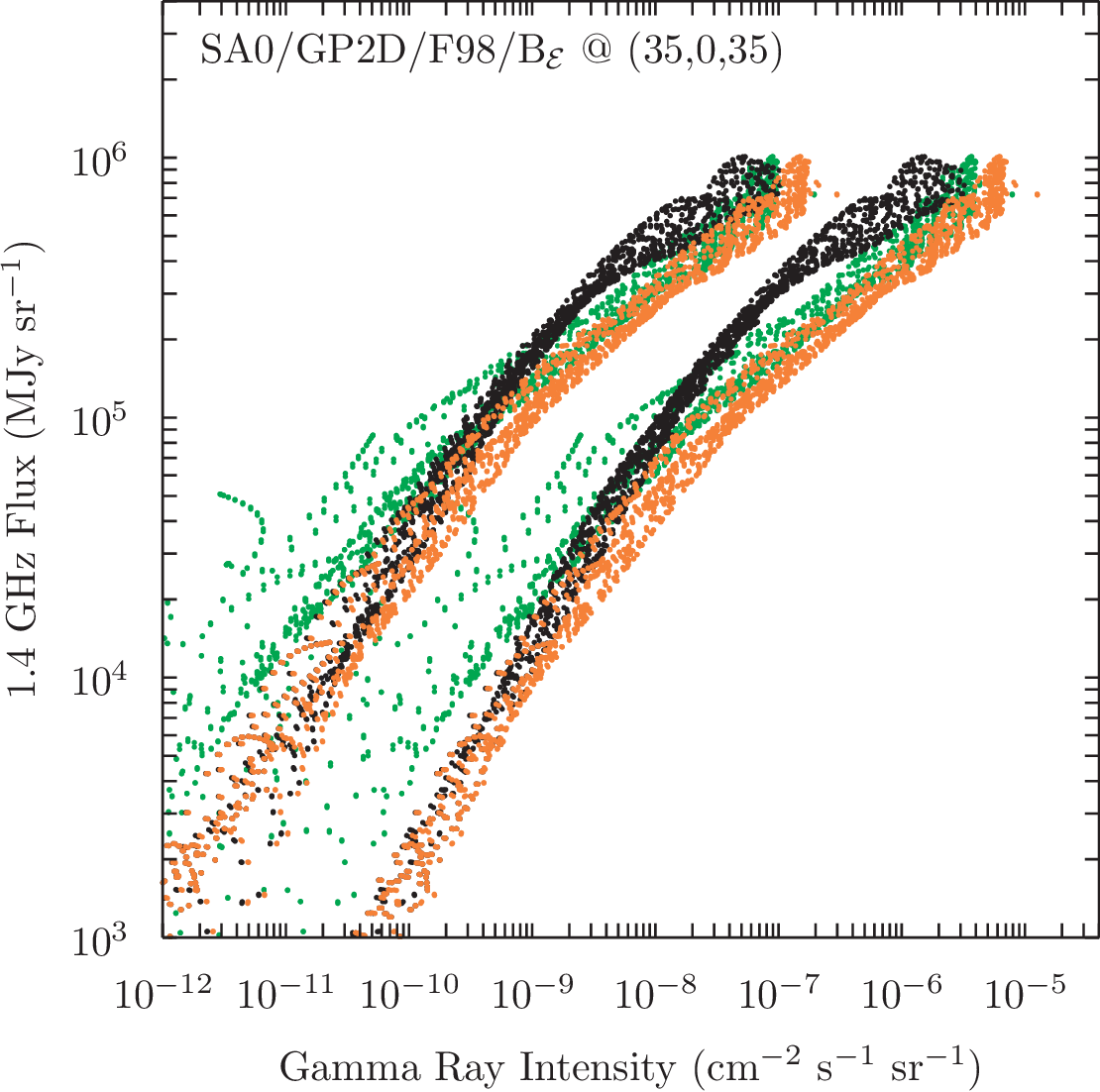}\hfill 
  \includegraphics[height=0.47\textwidth]{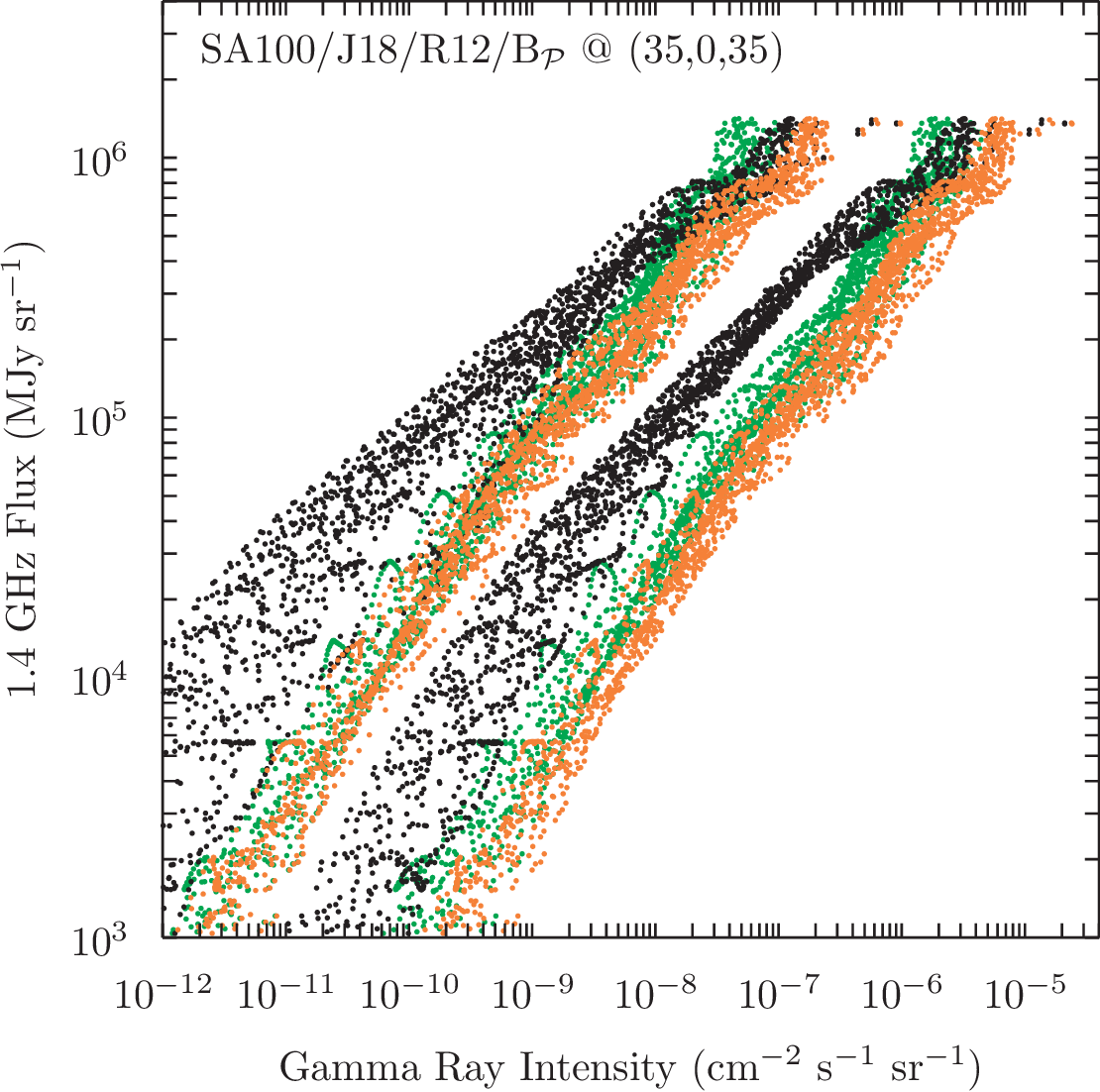}\hfill \\
  \includegraphics[height=0.47\textwidth]{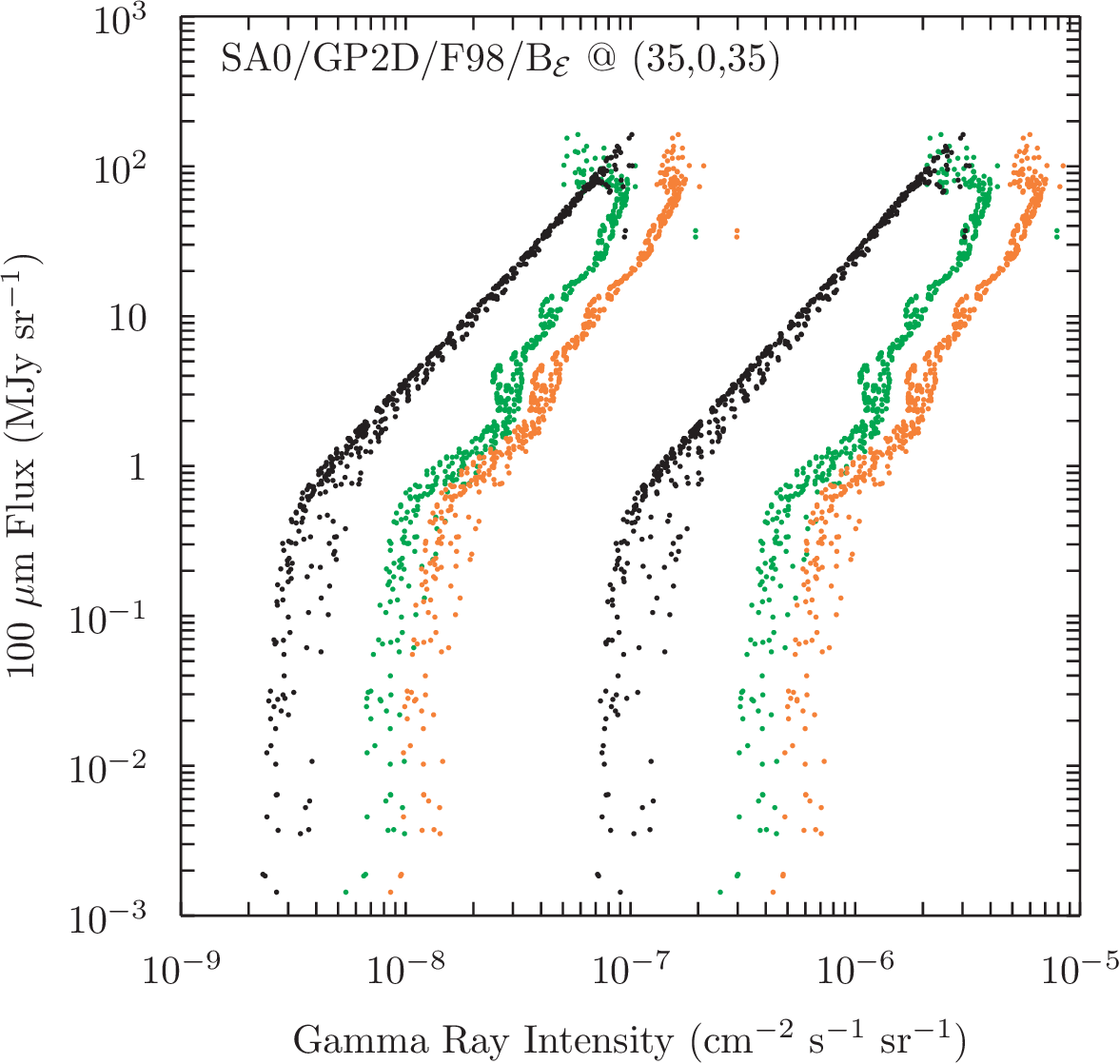}\hfill 
  \includegraphics[height=0.47\textwidth]{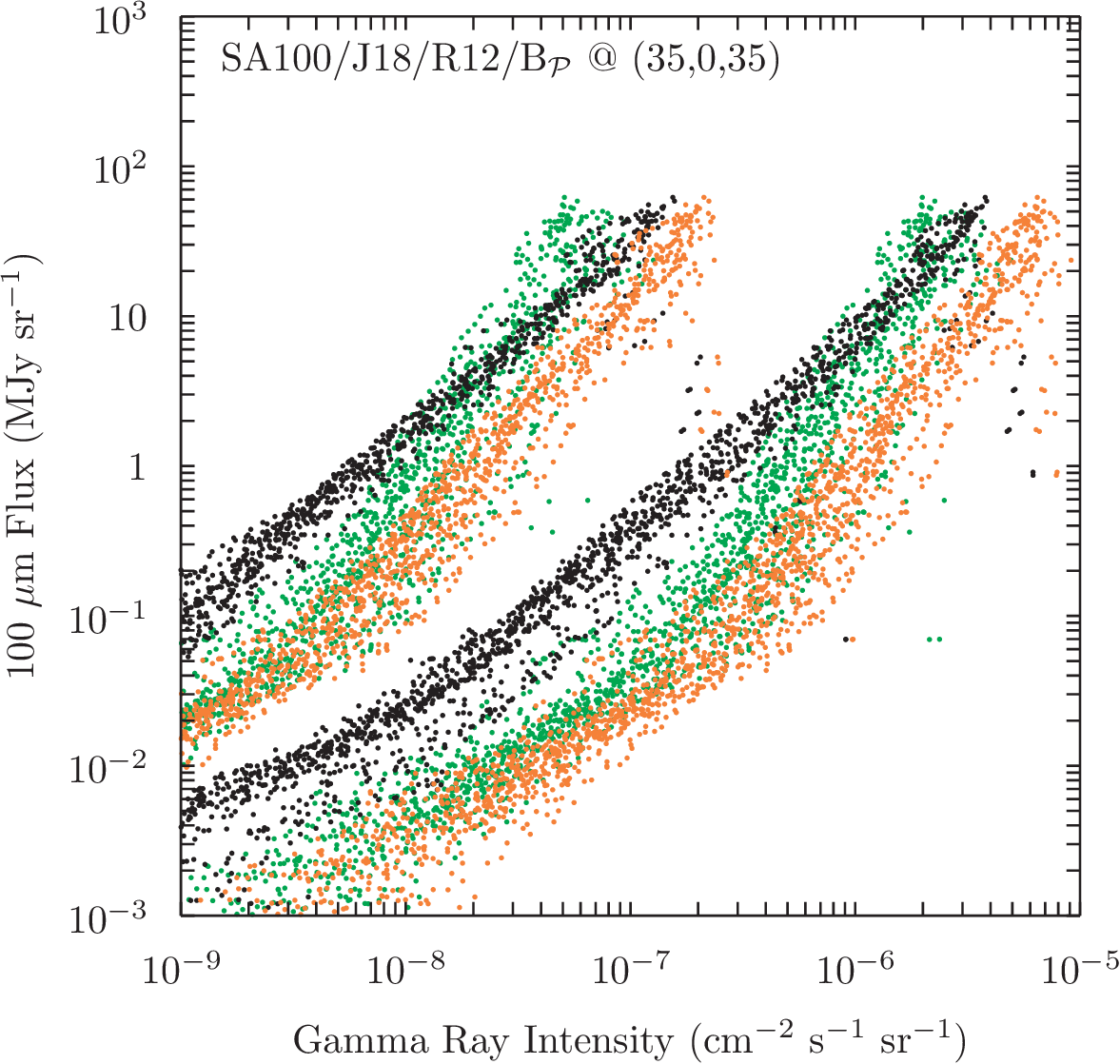}\hfill
  \caption{Correlation plots of (top) 1.4~GHz RC and (bottom) 100~$\mu$m IR flux vs. \gray{} intensity for the SA0-GP2D-F98-B$_\mathcal{E}$ (left) and SA100-J18-R12-B$_\mathcal{P}$ (right) modelling configurations for an observer located at 50~kpc at an inclination of 45$^\circ$ from the Galactic plane looking toward the GC of the MW.
For the \gray{} intensities, the correlations are separated into production processes: gas/$\pi^0$-decay, green; ISRF/Compton, black; total, orange. For each of these, the left point cloud corresponds to 10--100~GeV \gray{s} and the right to 1--10~GeV \gray{s}. \label{fig:ext_obs50kpc_hp6_45deg} }
\end{figure*}

The top panels show the 1.4~GHz RC--\gray{} correlation, and the bottom panels the 100~$\mu$m IR--\gray{} correlation.
For the RC--\gray{} correlation, a quasi-linear trend is observed for both model configurations.
This scaling is primarily driven by the large-scale radial decline of the CR sources and ISM components, the gas, ISRF, and magnetic fields, from the GC.
This gradient is a common feature of both the simple 2D and complex 3D models, as both are constrained by the same observational data for the MW.
While CRs propagate from their sources and their distribution is smoothed, the face-on LOS integration averages over the local structures of each model.
This process makes the resulting integrated intensities a tracer of the common radial gradient.

The IR-\gray{} correlation also shows a quasi-linear trend but is more sensitive to the details of the ISM component distributions.
For the simple axisymmetric model (bottom left), this sensitivity manifests as distinct numerical artifacts.
At high intensities, corresponding to regions within the Solar circle, prominent undulations for the correlation arise from the interpolation of the binned gas profile used for its density distribution over the inner galaxy.
Outside this the gas density follows a more regular exponential fall-off.
The 2D symmetry of the model when viewed from above at the limited resolution shown produces clusters along the correlation as the intensities decrease.
Beyond these artifacts, the model also shows physically motivated deviations at the intensity extremes: at high intensities, the correlation weakens as the IR flux is suppressed by the central hole in the F98 dust model while the \gray{} flux continues to rise.
At low intensities, a cutoff appears where the IR-emitting dust distribution terminates, while the more extended CR and gas distributions still produce faint \gray{} emission.
For the more complex 3D model (bottom right), these geometry-dependent artifacts are replaced by an increased, more realistic scatter.
This reflects the reduced point-to-point correlation between the structured gas in the J18 distribution and the more diffuse CR distribution.

Essentially, the 3D configuration in the right-hand panels produces a cloud of points corresponding to the arm-interarm contrast: CR sources and gas are concentrated in the arms, while the GMF and ISRF distributions are smooth and extend into the interarm region. The horizontal series of points visible in the high-intensity part of the 1.4 GHz RC--\gray{} correlation correspond to nearly constant radio intensity, while the spread in \gray{} intensity arises from pixels that span the arm-interarm transition. The features at the low-intensity end arise because the CR-gas and GMF distributions have different effective radial truncation lengths. At higher angular resolution, these features would be resolved into a continuous cloud, as can be seen in the following figures produced at full resolution.

The sensitivity of the RC--\gray{} correlation when the individual Galactic components are varied is shown in Fig.~\ref{fig:ext_obs50kpc_hp9_radiogamma}.
The SA0 source model is held constant and the target ISM components are changed, with each configuration normalised to the same CR data.
The correlation plots are made for the same observer location as Fig.~\ref{fig:ext_obs50kpc_hp6} but using the full resolution $N_{\rm side}$=512 images without degradation.

The most significant effect on the dominant hadronic component comes from the choice of gas model.
When the GP2D distribution is replaced with the 3D J18 one (bottom left panel), the unphysical features in the gas/$\pi^0$-decay component and total intensity stemming from the former's binned interpolation are replaced by a broader, more realistic scatter.
This physical scatter is a direct consequence of the imperfect spatial correlation on sub-kpc scales between the non-axisymmetric gas structures in the J18 distribution and the more smoothly distributed CR population from the axisymmetric SA0 source model.
Changing the magnetic field structure, in turn, directly impacts the radio emission.
Altering the GMF from a simple random field to one including a coherent component (bottom right panel) introduces stronger spatial variations in the synchrotron emissivity.
This leads to greater variation in the radio intensity for a given CR electron density, thereby increasing the scatter in the RC--\gray{} correlation.
The choice of ISRF model (top right panel) modifies the sub-dominant IC \gray{} component.
The R12 model's higher optical photon density in the inner Galaxy boosts the IC flux in that region, altering the high-intensity end of the total \gray{} emission and thus subtly reshaping the correlation.

The IR--\gray{} correlation, shown in Fig.~\ref{fig:ext_obs50kpc_hp9_irgamma}, is particularly sensitive to the assumed dust and stellar distributions, which are coupled within the ISRF models.
The difference between the F98 and R12 models (top panels) illustrates this at both the high- and low-intensity ends of the correlation.
In the inner Galaxy, the R12 model has a larger central hole in its dust distribution.
This leads to a lower IR flux but a higher optical photon density, which in turn boosts the Compton \gray{} emission as CR electrons upscatter these optical photons most strongly for the 1--100~GeV \gray{} band.
It manifests as a sharp drop in the correlation of the IR flux at high \gray{} intensity.

Conversely, the F98 model, with its greater dust mass in the same region, produces a higher IR flux, altering the slope and termination point of the IR--\gray{} correlation at the highest intensities.
The F98 model also features a sharp radial truncation of its stellar and dust discs, which causes an abrupt cutoff in the IR flux at large galactocentric radii and truncates the correlation at low intensities.
The R12 model, which has a more extended distribution, does not exhibit this sharp cutoff, allowing the correlation to continue to lower fluxes.
The gas distribution (bottom left panel) affects the IR--\gray{} correlation in the same manner as the RC--\gray{} case; replacing the GP2D distribution with the J18 one introduces physical scatter due to the imperfect gas-CR spatial correlation.
As expected, altering the GMF model has essentially no effect on this correlation, because neither the IR nor \gray{} emissions are directly dependent on its distribution.

The CR source distribution significantly affects the correlation, as shown in Fig.~\ref{fig:ext_obs50kpc_hp6_srcvar}.
For these figures we return to using the degraded resolution $N_{\rm side}$=128 maps as a convenience, because it is easier to distinguish the different point clouds.
A key diagnostic feature is the behaviour of the hadronic \gray{} component (green points) in the RC--\gray{} correlation.
The SA50 model exhibits a unique trend: at radio intensities below $\sim$3--4$\times10^4$~MJy~sr$^{-1}$, the hadronic \gray{} flux plateaus while the radio flux continues to decrease.
This feature originates from the hybrid source morphology of the SA50 model.

This behaviour can be understood by comparing the CR populations in the inter-arm regions of the three models.
In the SA0 model, the smooth injection of both CR protons and electrons leads to morphologically similar particle distributions and a tight, linear correlation down to the lowest fluxes.

The SA100 model is different. With all CRs originating in the arms, the differing propagation and energy loss scales of protons and electrons become critical. Energetic electrons cool rapidly, so their density falls off sharply away from the arms. Protons propagate much further. This creates a spatially varying proton-to-electron ratio across the galaxy, where the ratio is lowest near the arms and highest in the far inter-arm regions. This does not break the correlation, but broadens it significantly. Pixels near the arms have high radio and high \gray{} flux. Pixels far from the arms have low flux in both bands. However, the wide range of proton-to-electron ratios at intermediate distances creates a large scatter around the mean trend, resulting in the broad, fan-like distribution seen in the plot.

\begin{figure*}[t!]
  \centering
  \includegraphics[height=0.47\textwidth]{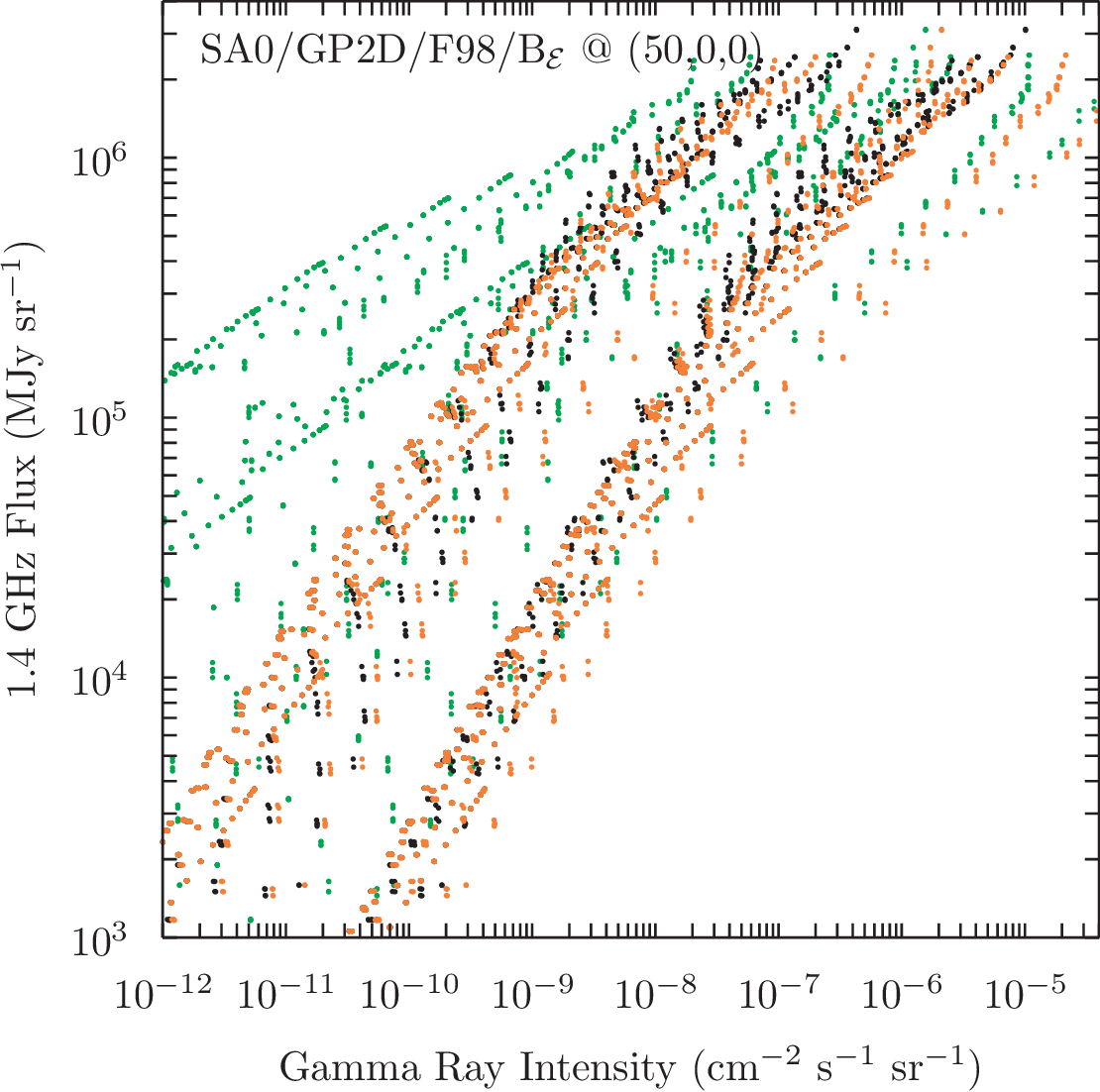}\hfill 
  \includegraphics[height=0.47\textwidth]{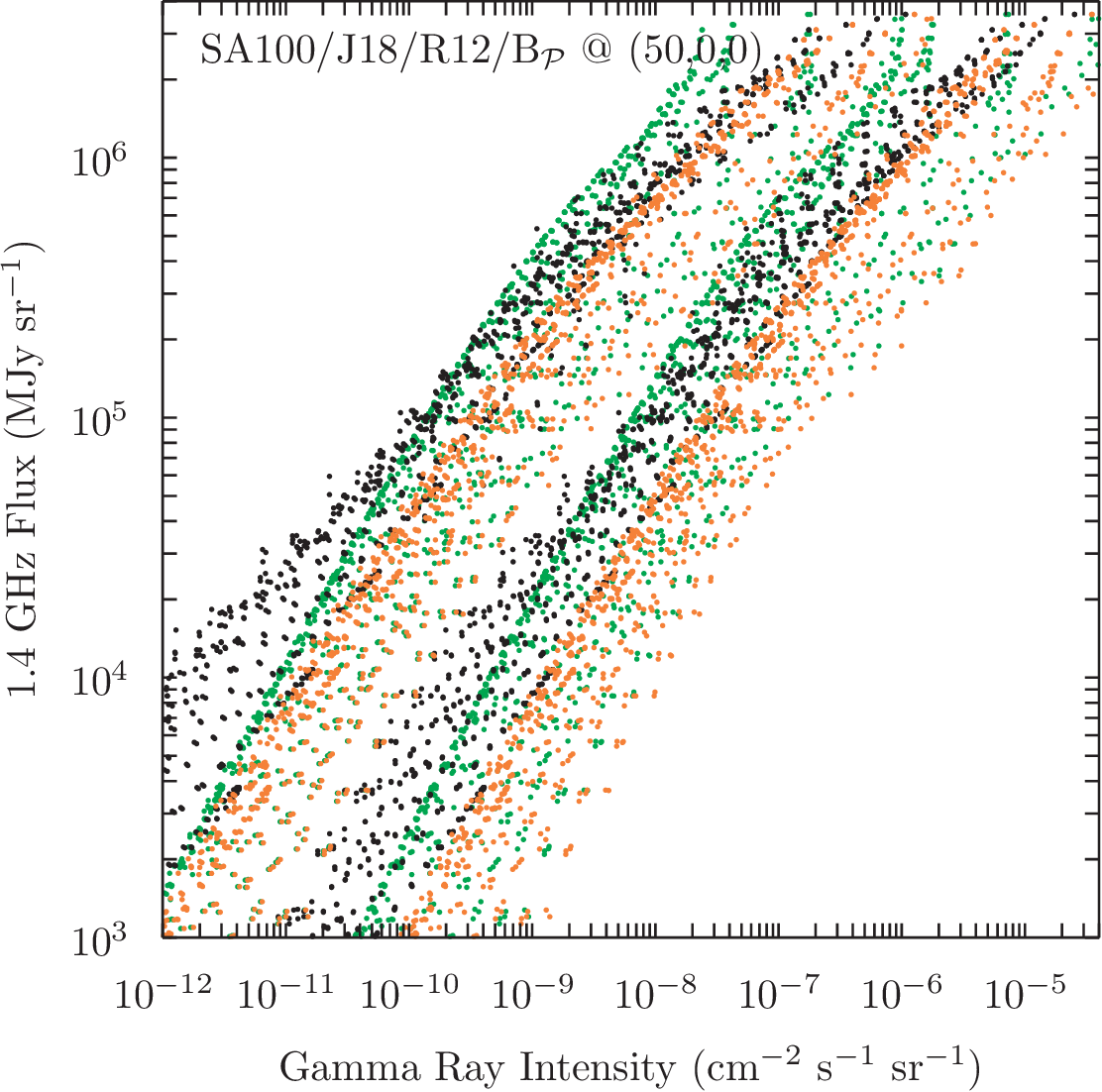}\hfill \\
  \includegraphics[height=0.47\textwidth]{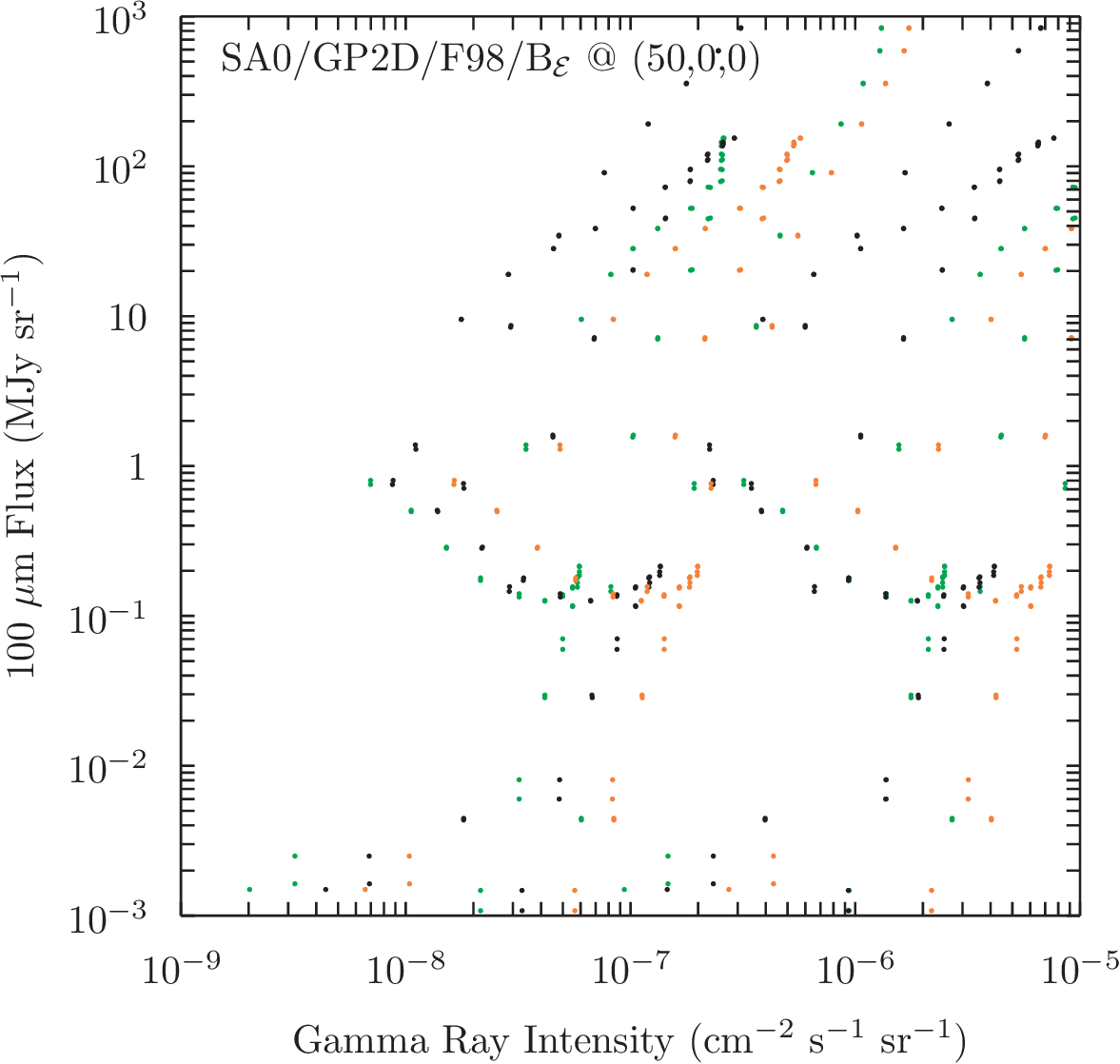}\hfill 
  \includegraphics[height=0.47\textwidth]{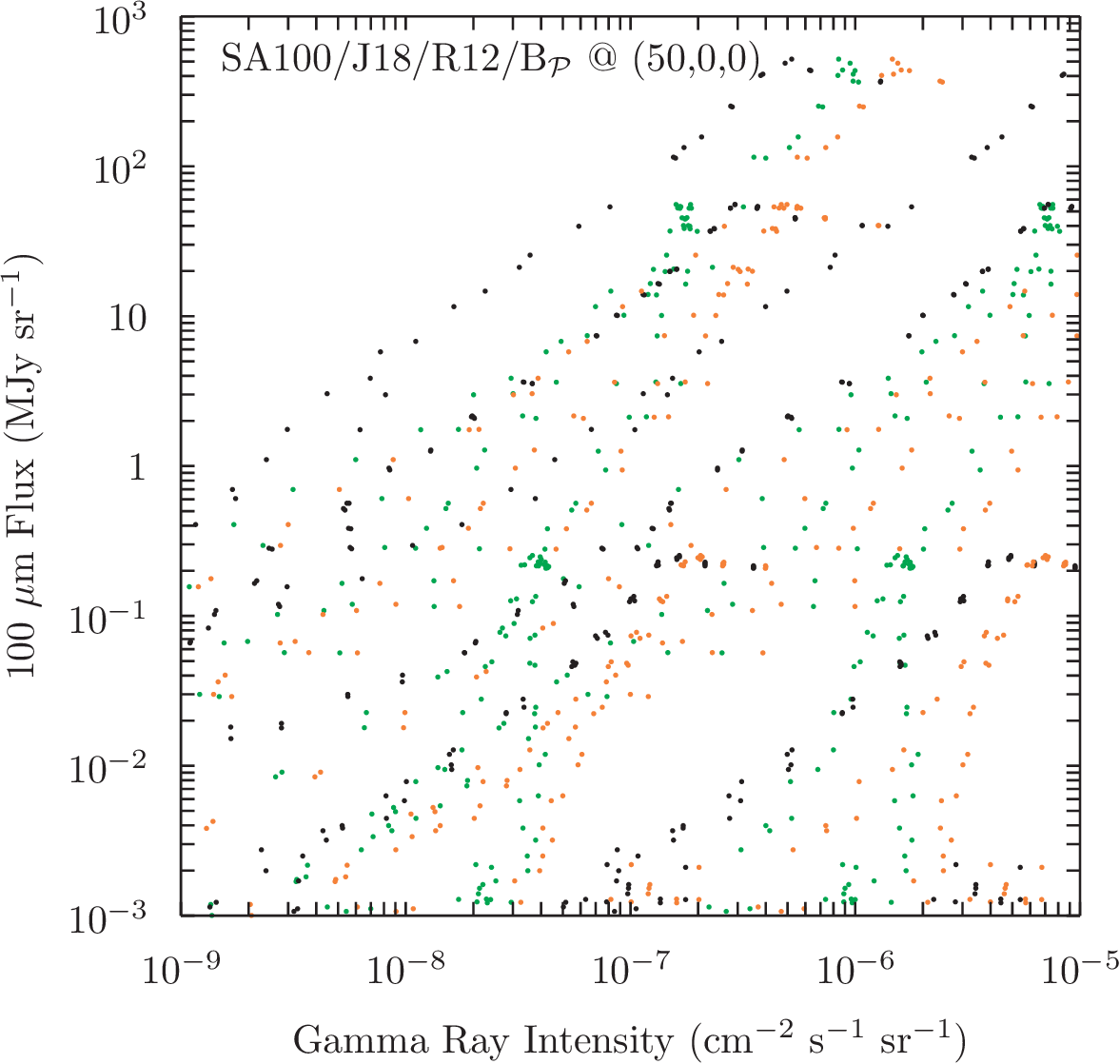}\hfill
  \caption{Correlation plots of (top) 1.4~GHz RC and (bottom) 100~$\mu$m IR flux vs. \gray{} intensity for the SA0-GP2D-F98-B$_\mathcal{E}$ (left) and SA100-J18-R12-B$_\mathcal{P}$ (right) modelling configurations for an observer located at 50~kpc at an inclination of 90$^\circ$ (edge-on) looking toward the GC of the MW.
For the \gray{} intensities, the correlations are separated into production processes: gas/$\pi^0$-decay, green; ISRF/Compton, black; total, orange. For each of these, the left point cloud corresponds to 10--100~GeV \gray{s} and the right to 1--10~GeV \gray{s}. \label{fig:ext_obs50kpc_hp6_edge} }
\end{figure*}

\begin{figure*}[t!]
  \centering
  \includegraphics[height=0.3\textwidth]{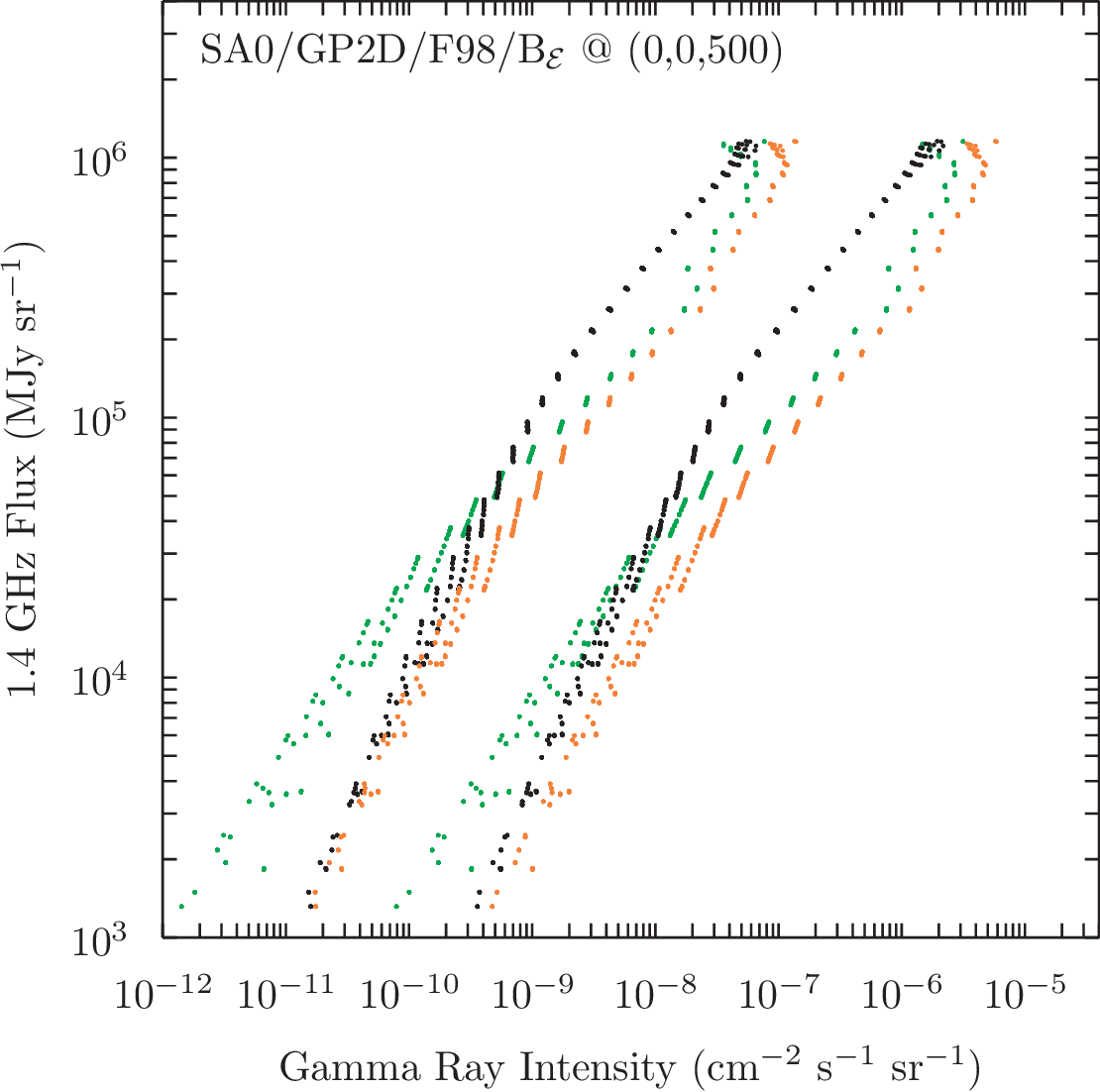}\hfill 
  \includegraphics[height=0.3\textwidth]{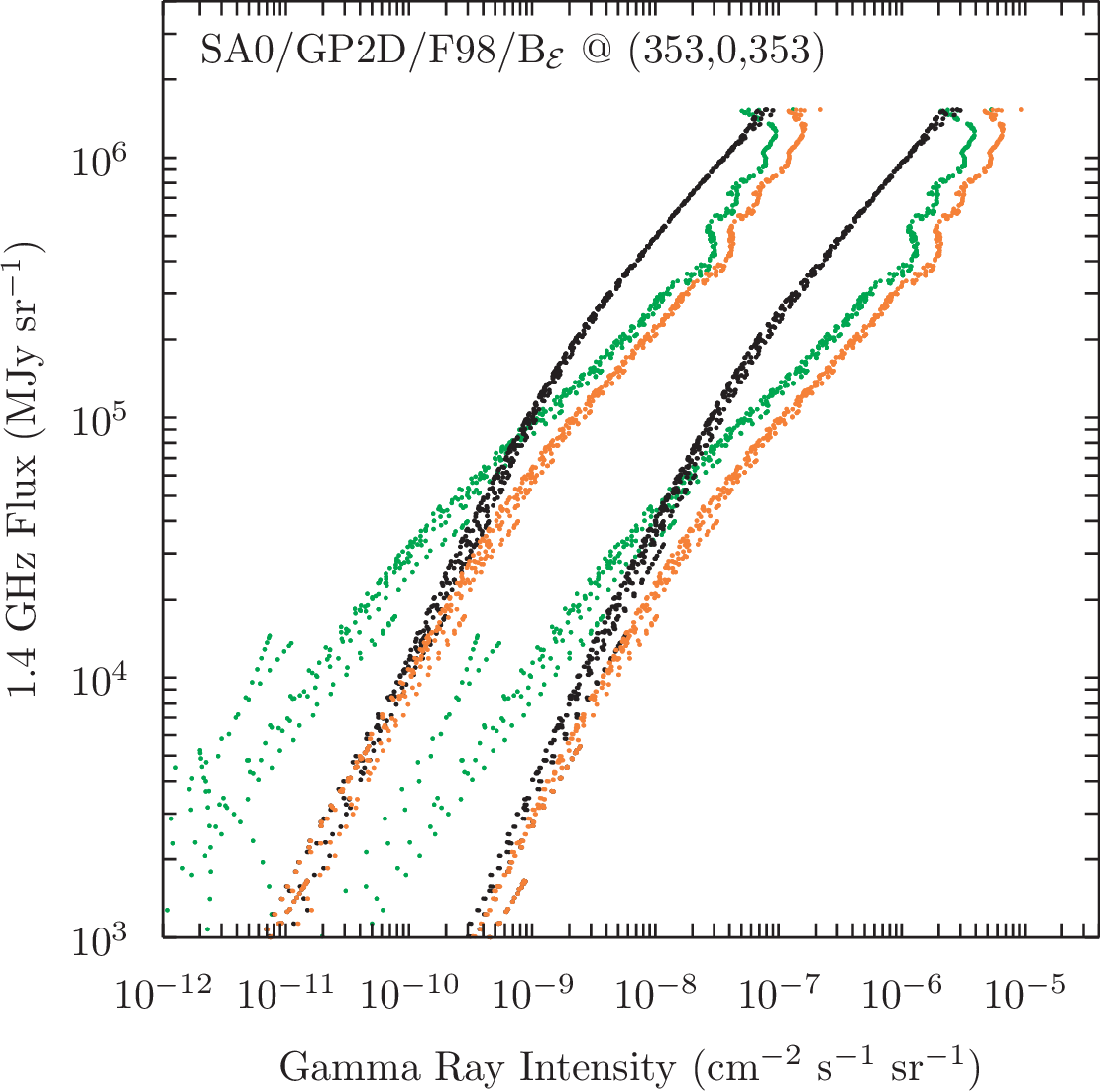} \hfill
  \includegraphics[height=0.3\textwidth]{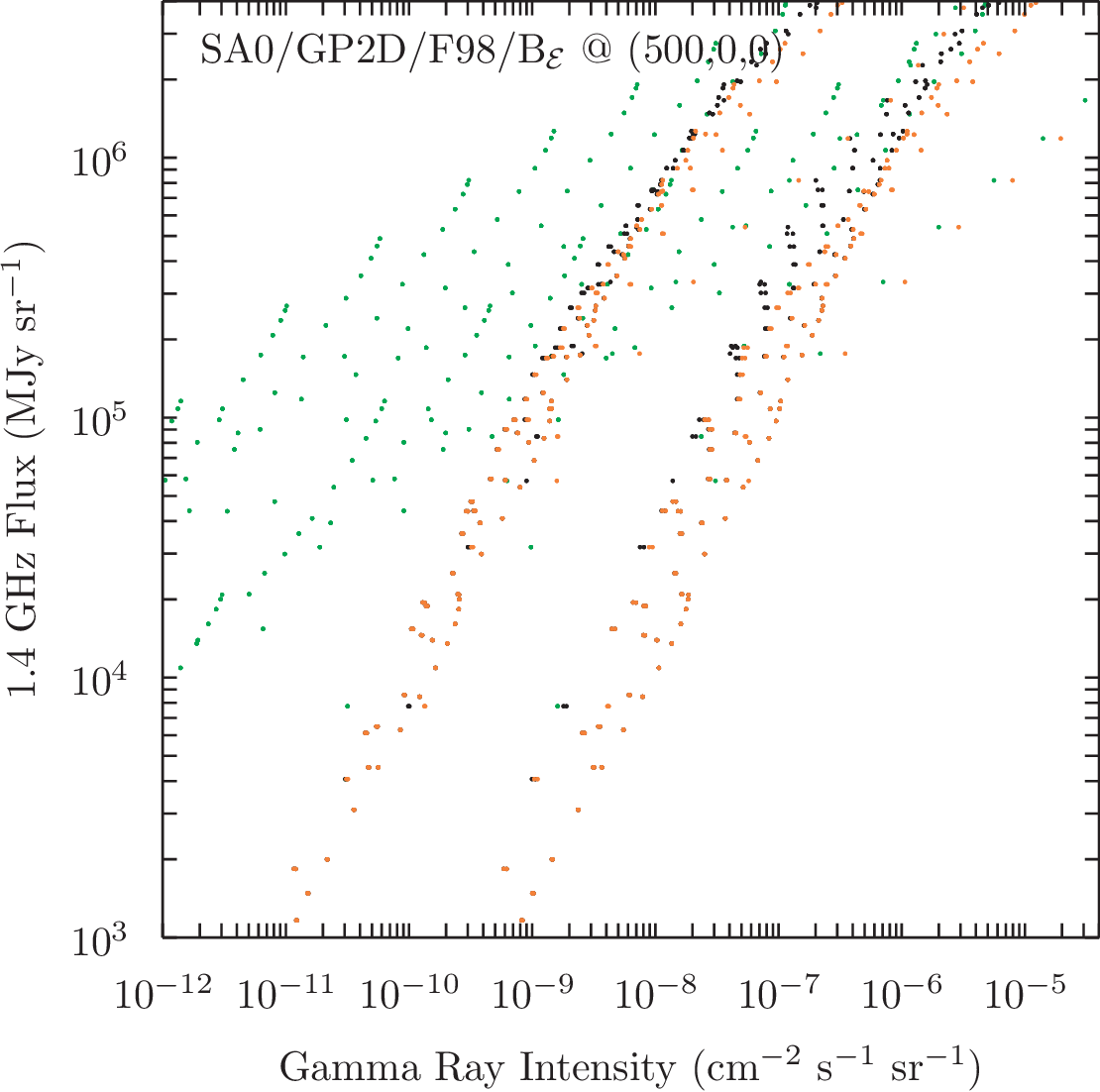} \hfill\\
  \includegraphics[height=0.3\textwidth]{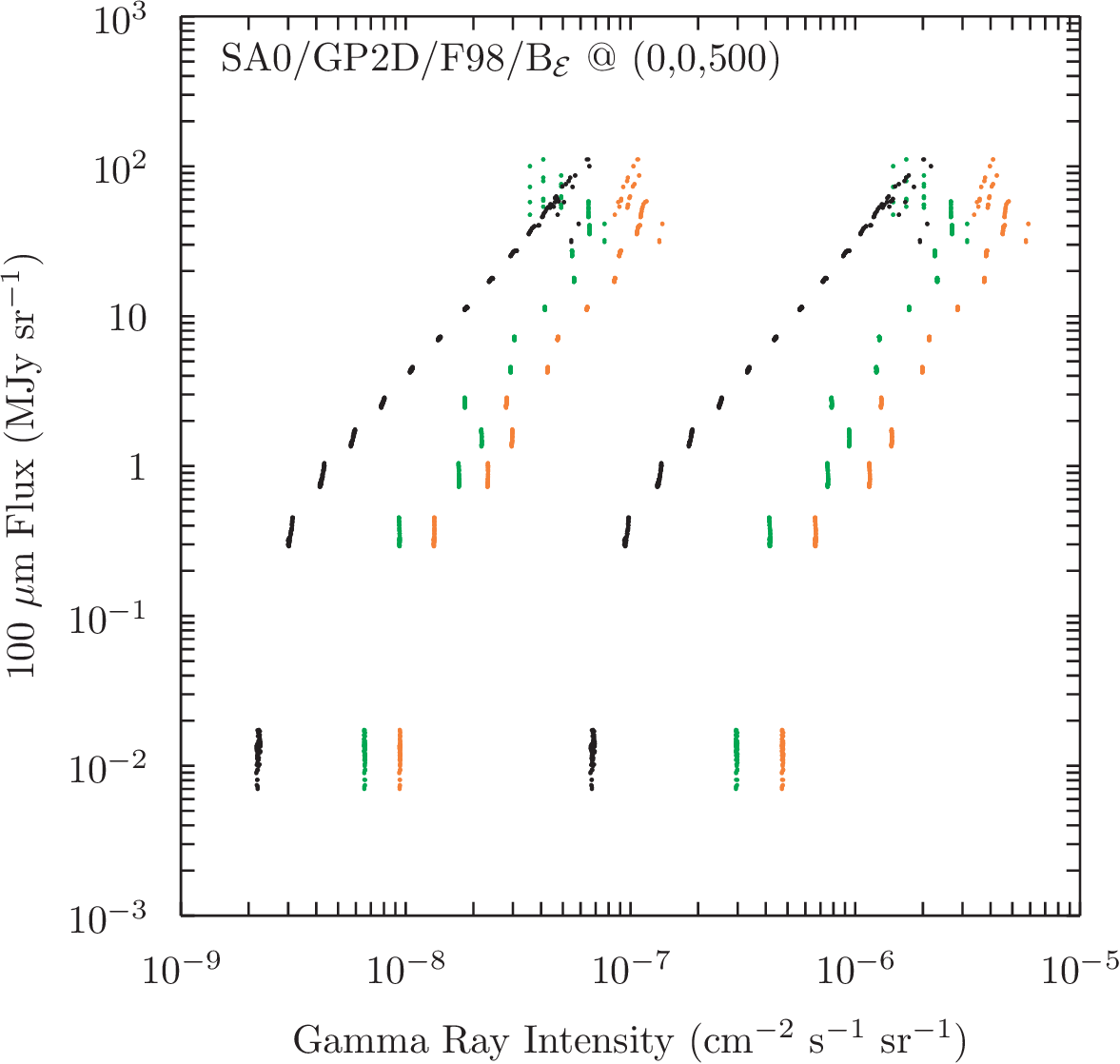}\hfill 
  \includegraphics[height=0.3\textwidth]{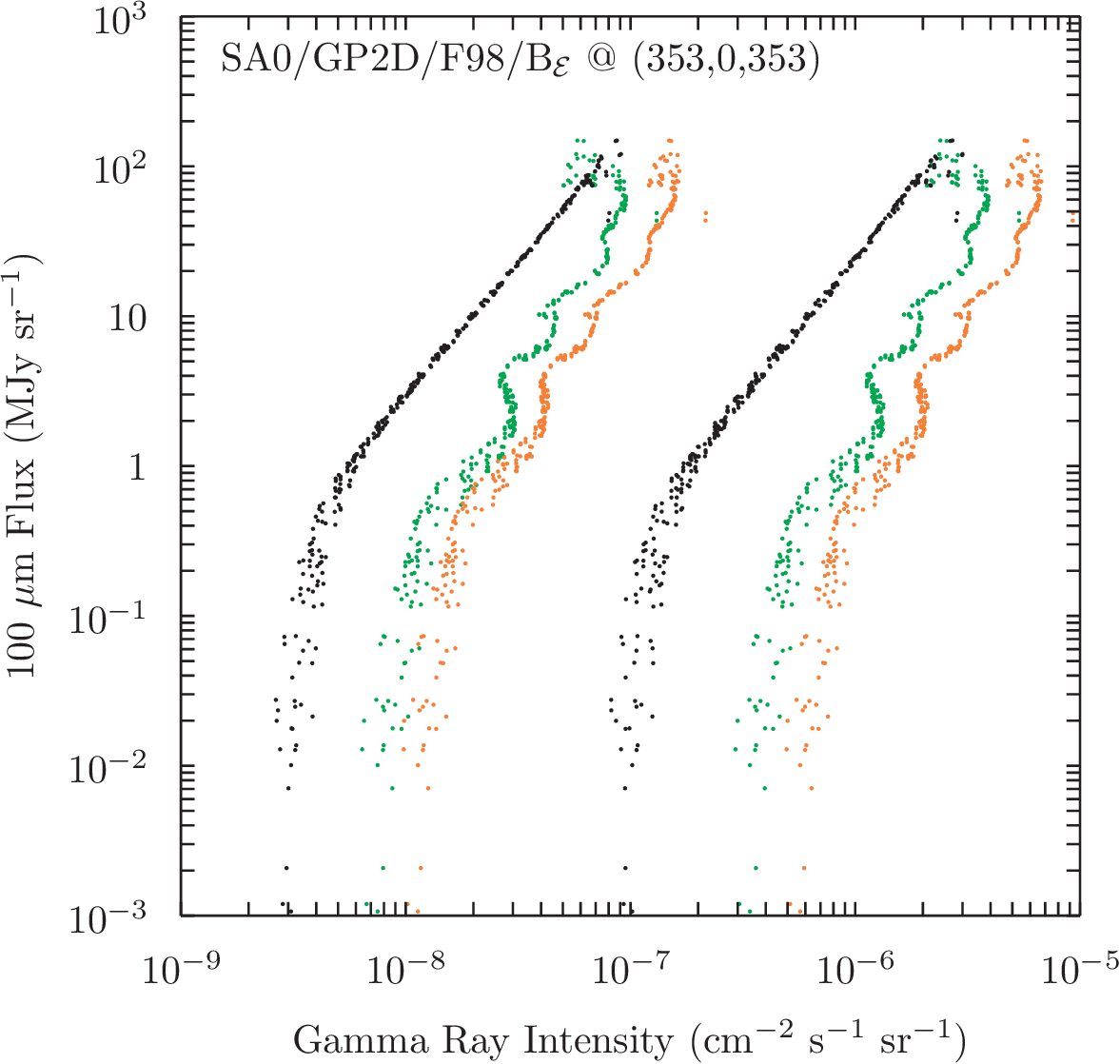} \hfill
  \includegraphics[height=0.3\textwidth]{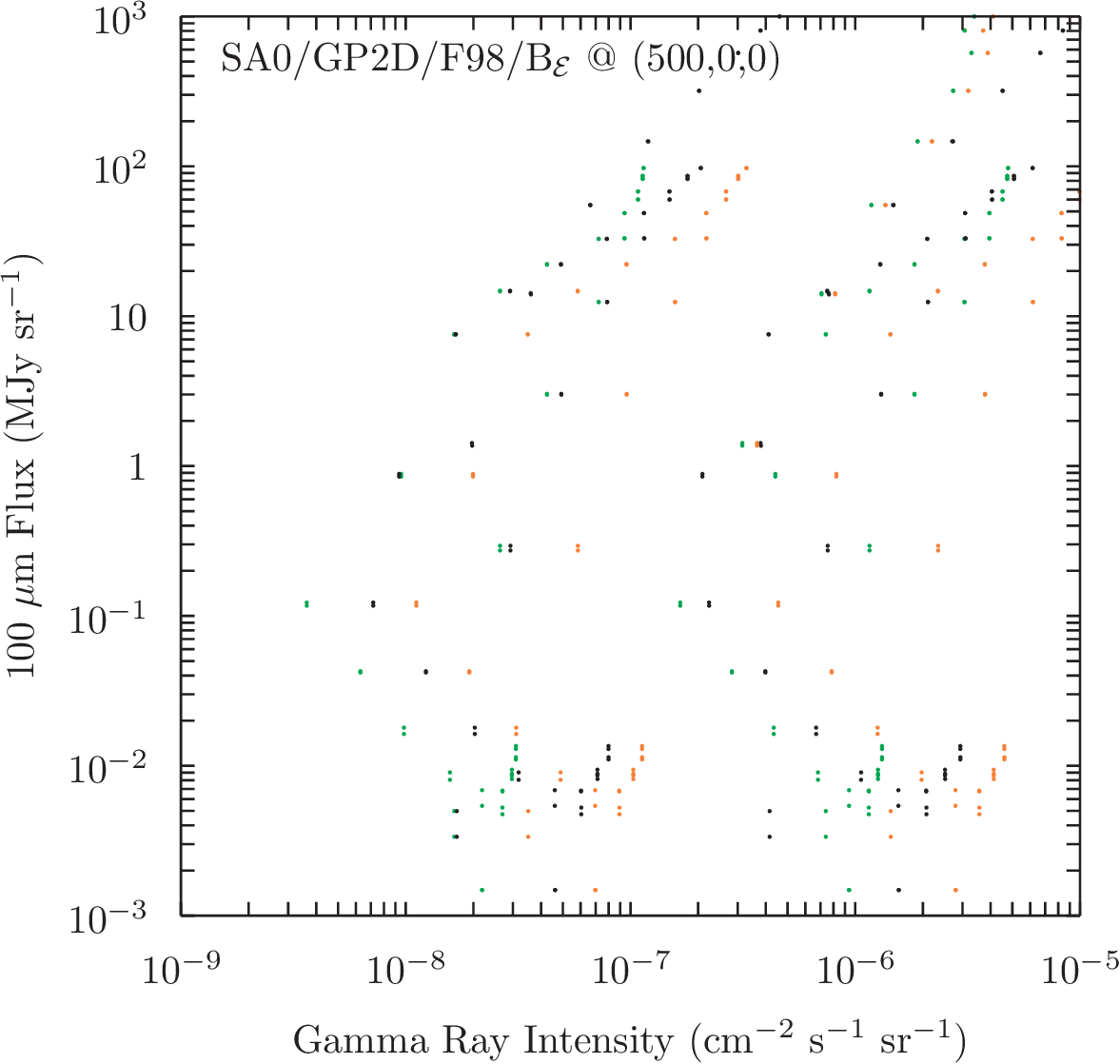} \hfill
  \caption{Correlation plots of (top) 1.4~GHZ flux vs.\ \gray{} intensity for
    the SA0-GP2D-F98-B$_\mathcal{E}$ and (bottom) 100$~\mu$m IR flux vs. \gray{} intensity for observer located 500~kpc from the MW centre looking face-down (left), at 45$^\circ$ inclination (middle), and edge-on (right).
    For the \gray{} intensities, the correlations are separated into production processes: gas/$\pi^0$-decay, green; ISRF/Compton, black; total, orange. For each of these, the left point cloud corresponds to 10--100~GeV \gray{s} and the right to 1--10~GeV \gray{s}.
    \label{fig:ext_obs500kpc_hp9_sa0} }
\end{figure*}

\begin{figure*}[t!]
  \centering
  \includegraphics[height=0.3\textwidth]{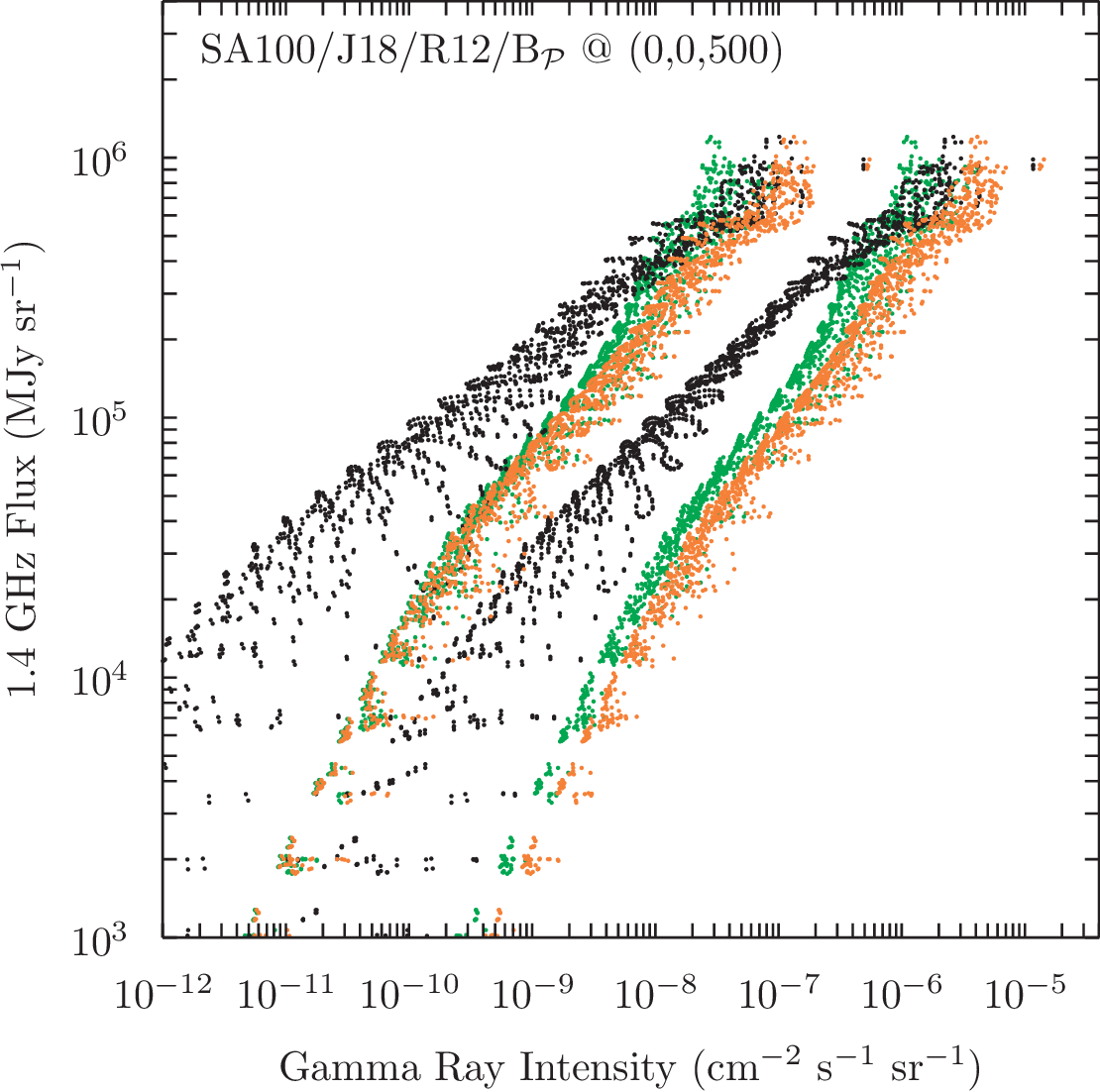}\hfill 
  \includegraphics[height=0.3\textwidth]{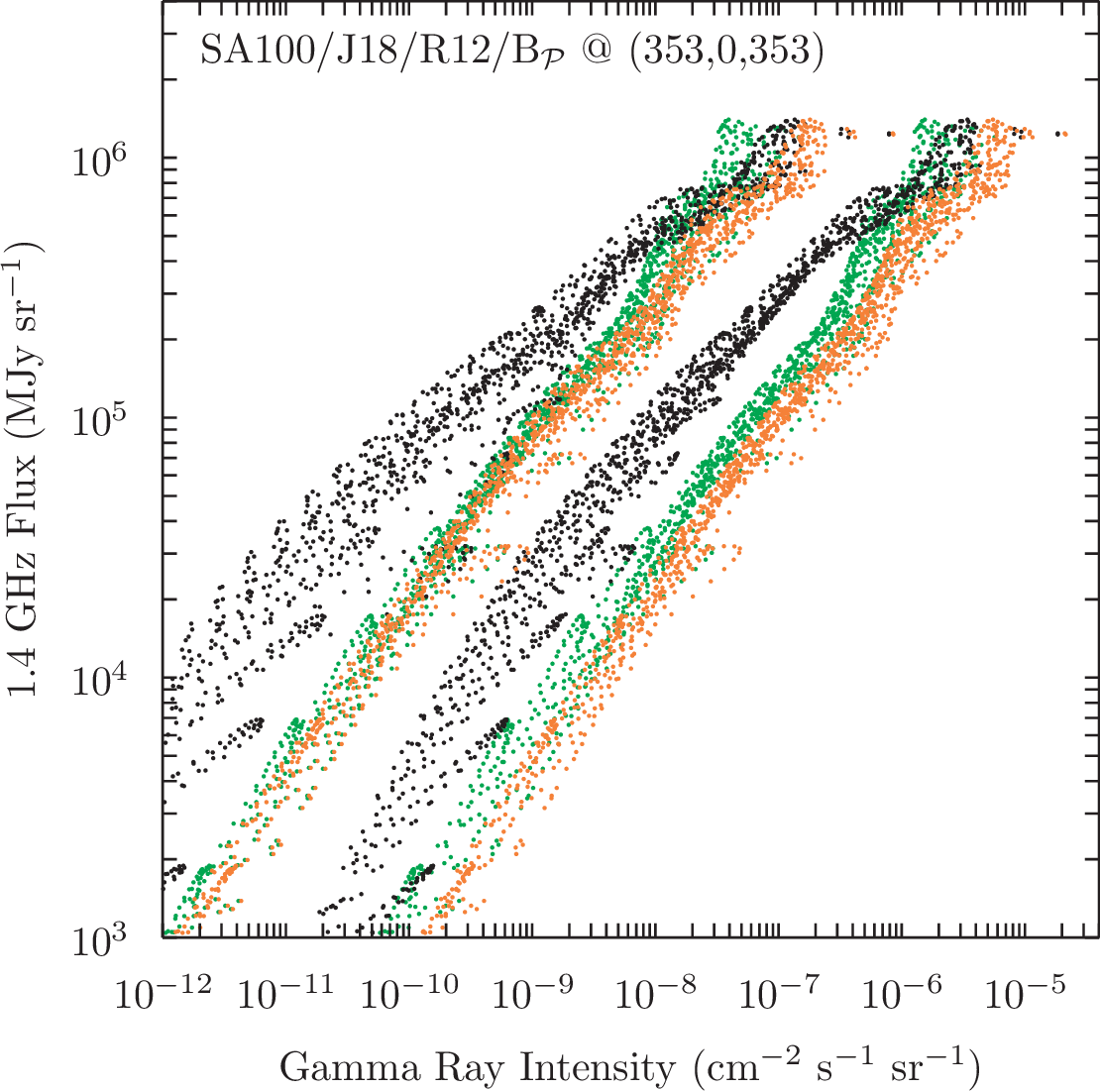} \hfill
  \includegraphics[height=0.3\textwidth]{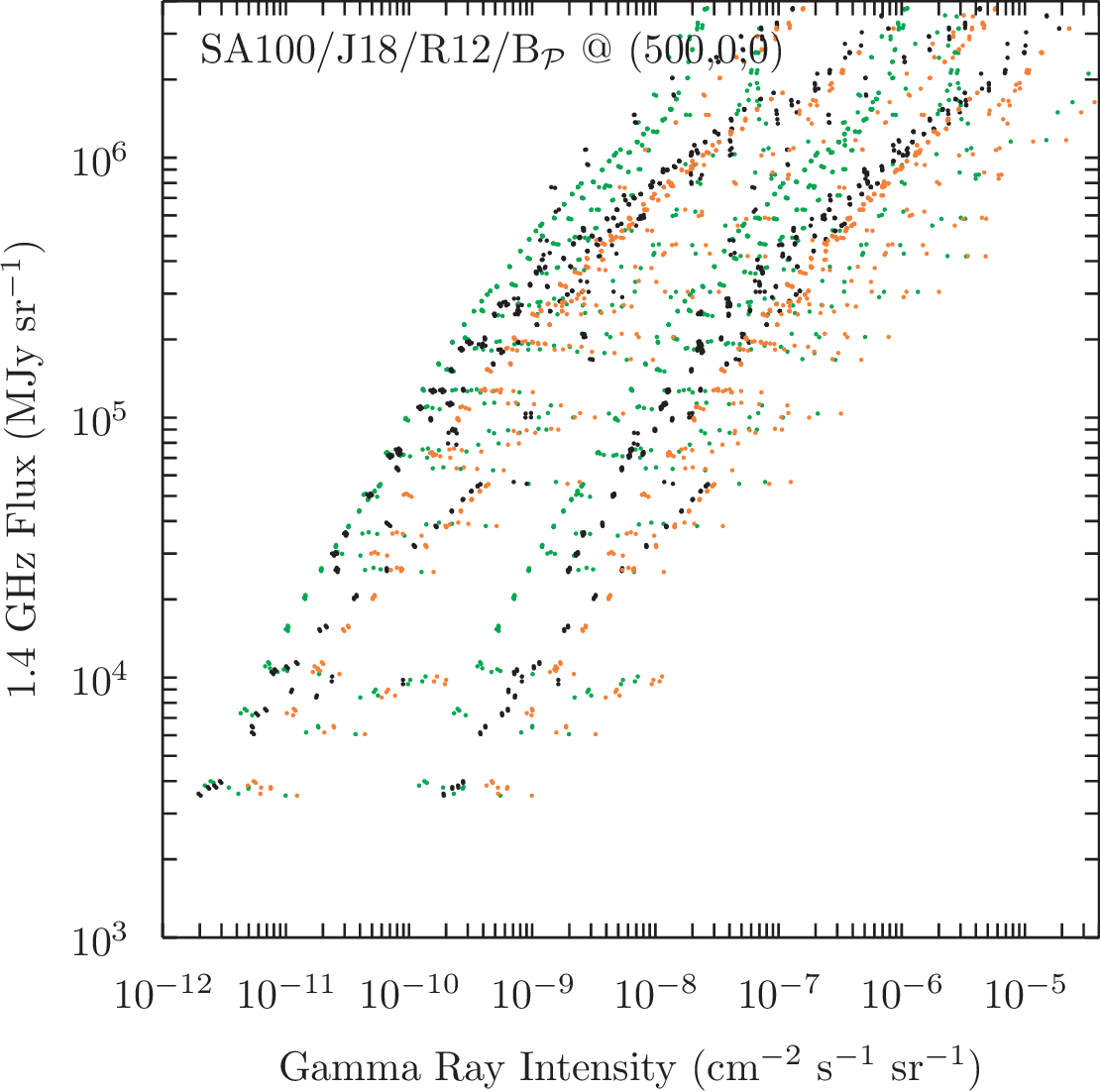} \hfill\\
  \includegraphics[height=0.3\textwidth]{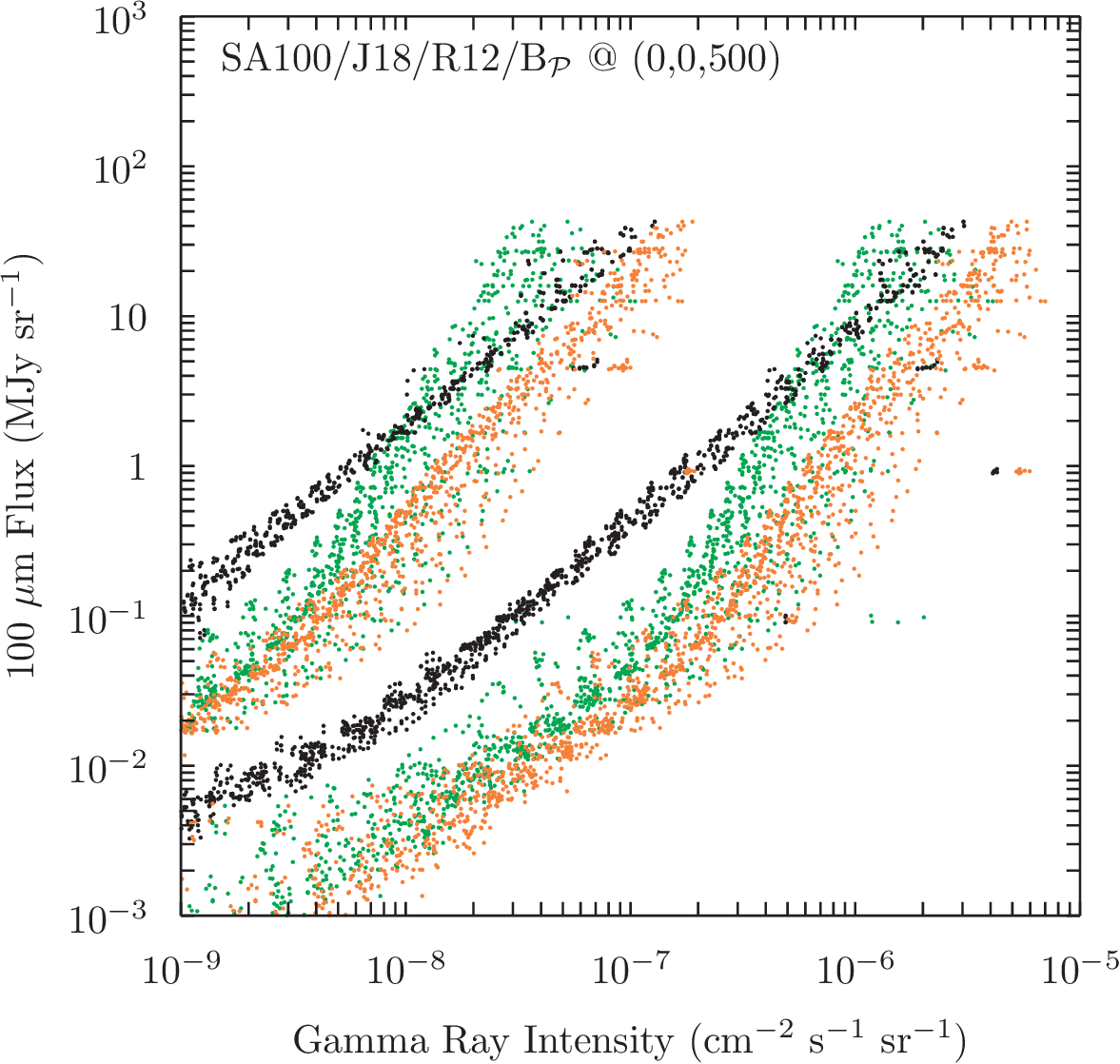}\hfill 
  \includegraphics[height=0.3\textwidth]{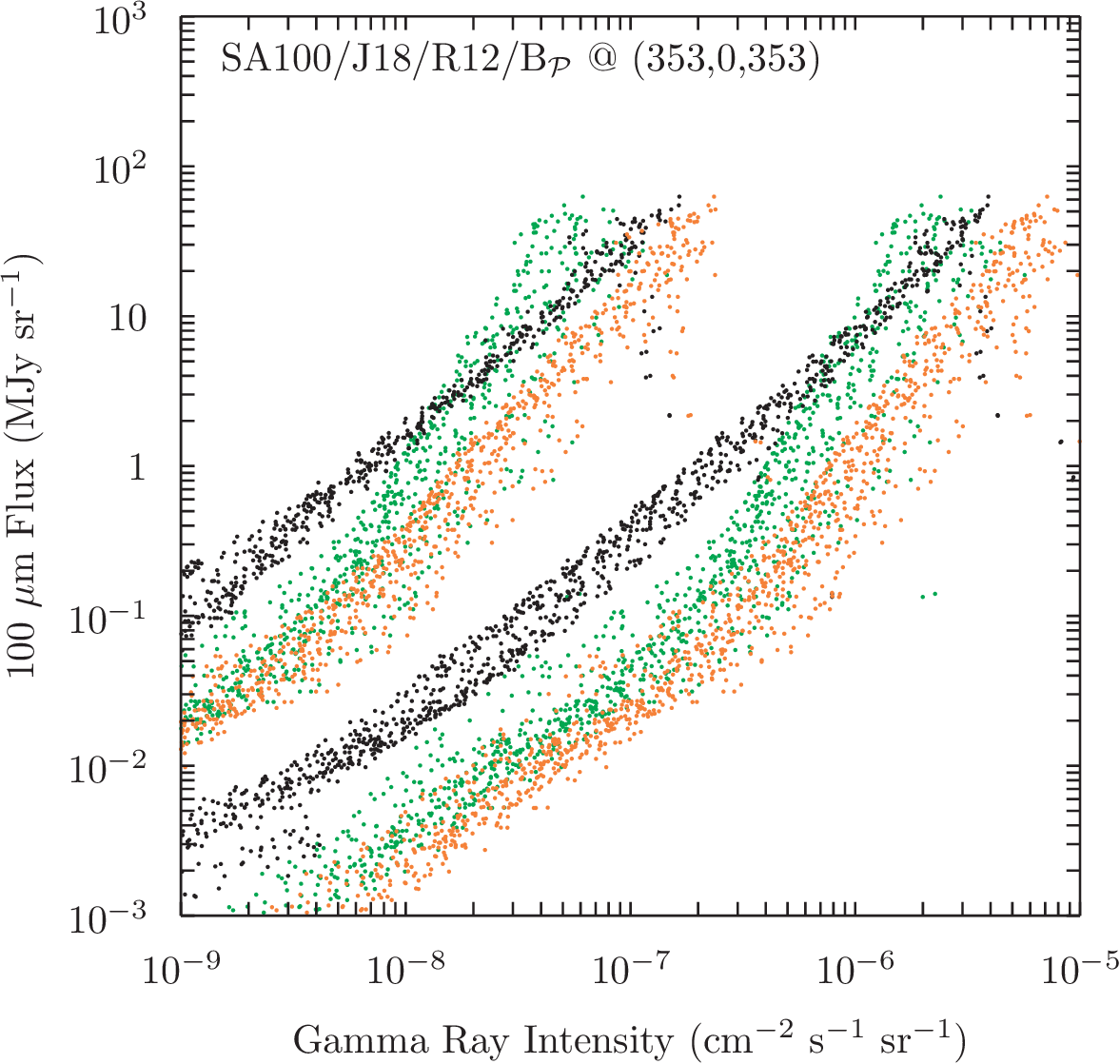} \hfill
  \includegraphics[height=0.3\textwidth]{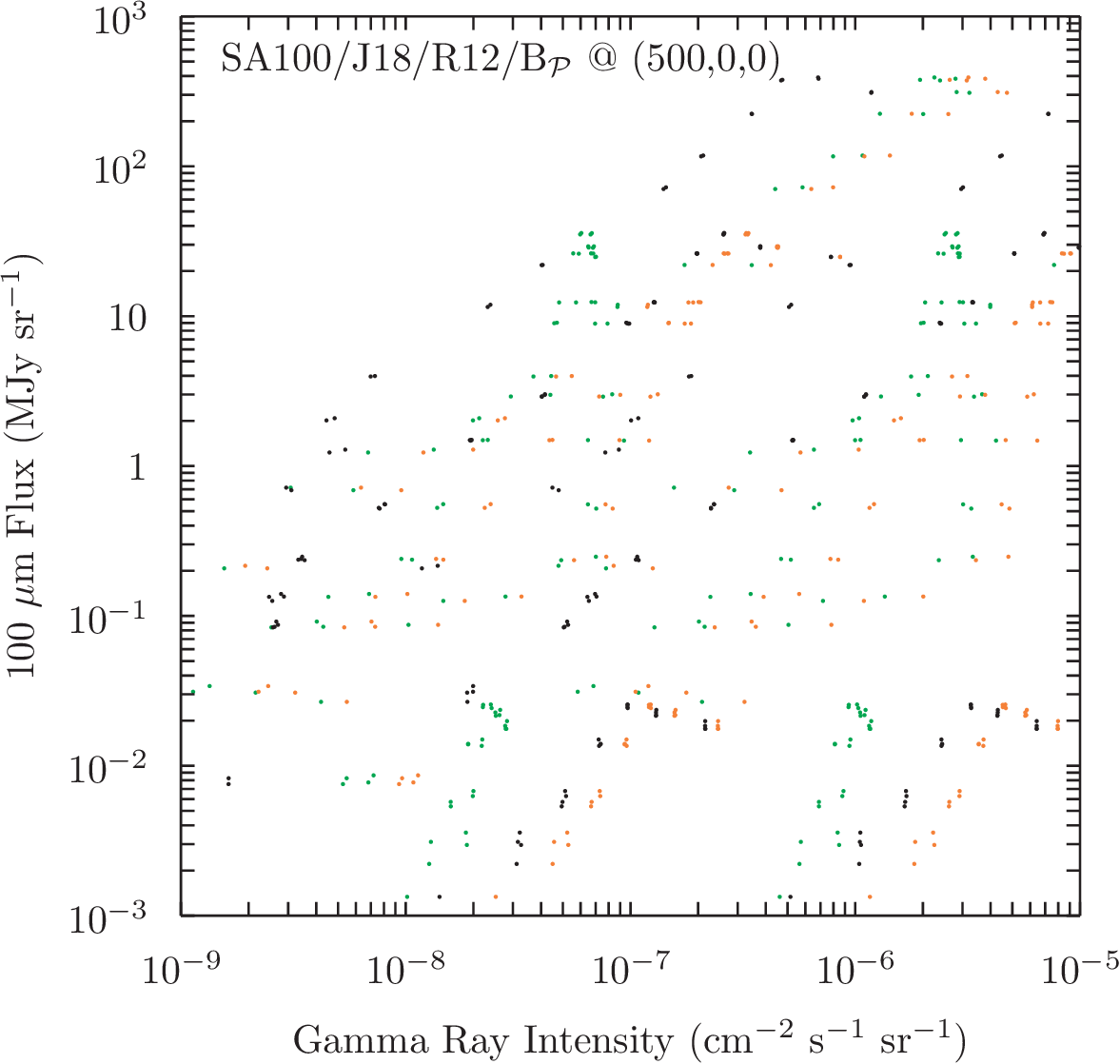} \hfill
  \caption{Correlation plots of (top) 1.4~GHZ flux vs.\ \gray{} intensity for
    the SA100-J18-R12-B$_\mathcal{P}$ and (bottom) 100$~\mu$m IR flux vs. \gray{} intensity for observer located 500~kpc from the MW centre looking face-down (left), at 45$^\circ$ inclination (middle), and edge-on (right).
    For the \gray{} intensities, the correlations are separated into production processes: gas/$\pi^0$-decay, green; ISRF/Compton, black; total, orange. For each of these, the left point cloud corresponds to 10--100~GeV \gray{s} and the right to 1--10~GeV \gray{s}.
    \label{fig:ext_obs500kpc_hp9_sa100} }
\end{figure*}

\begin{figure*}[t!]
  \centering
  \includegraphics[height=0.47\textwidth]{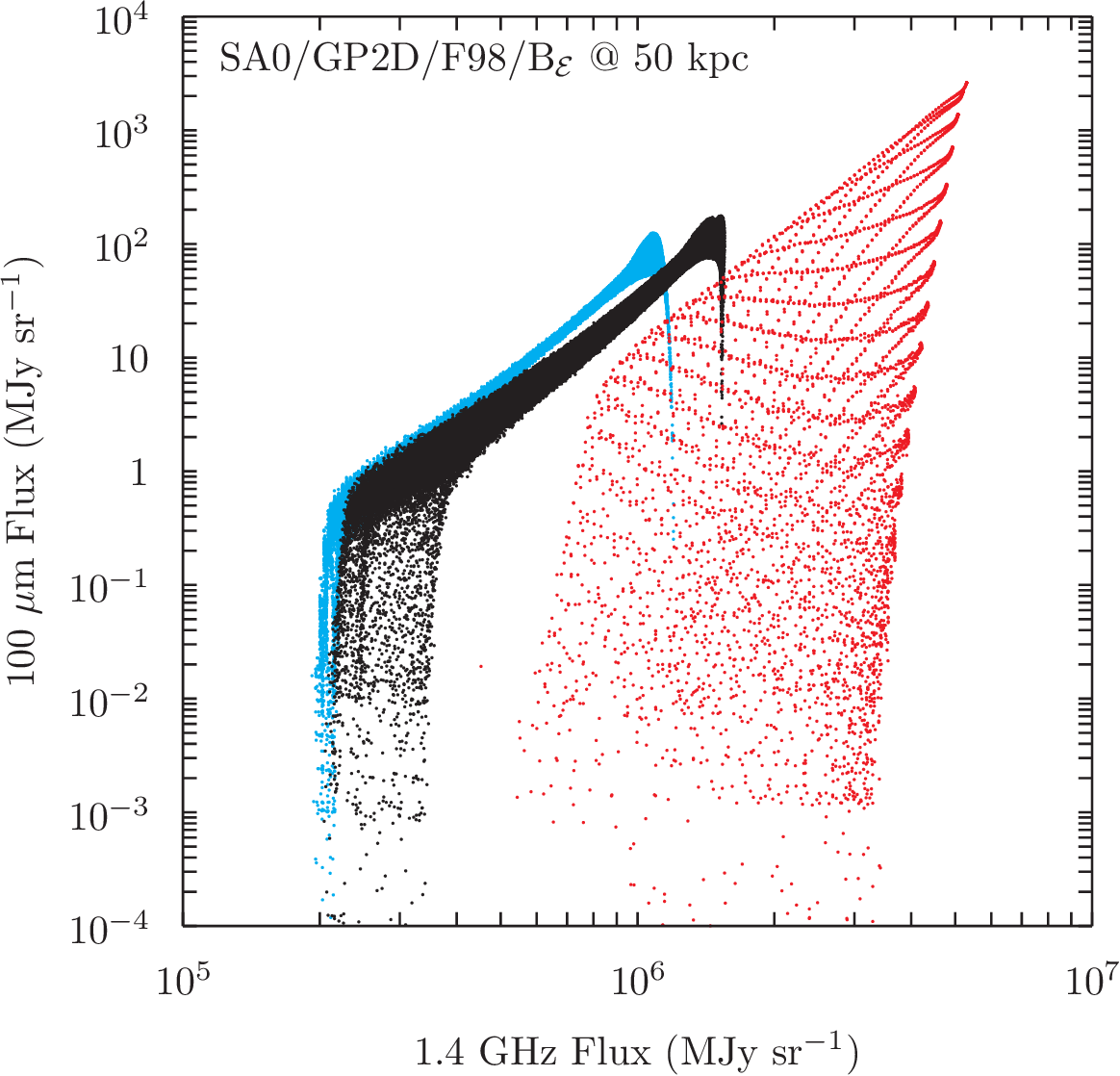}\hfill 
  \includegraphics[height=0.47\textwidth]{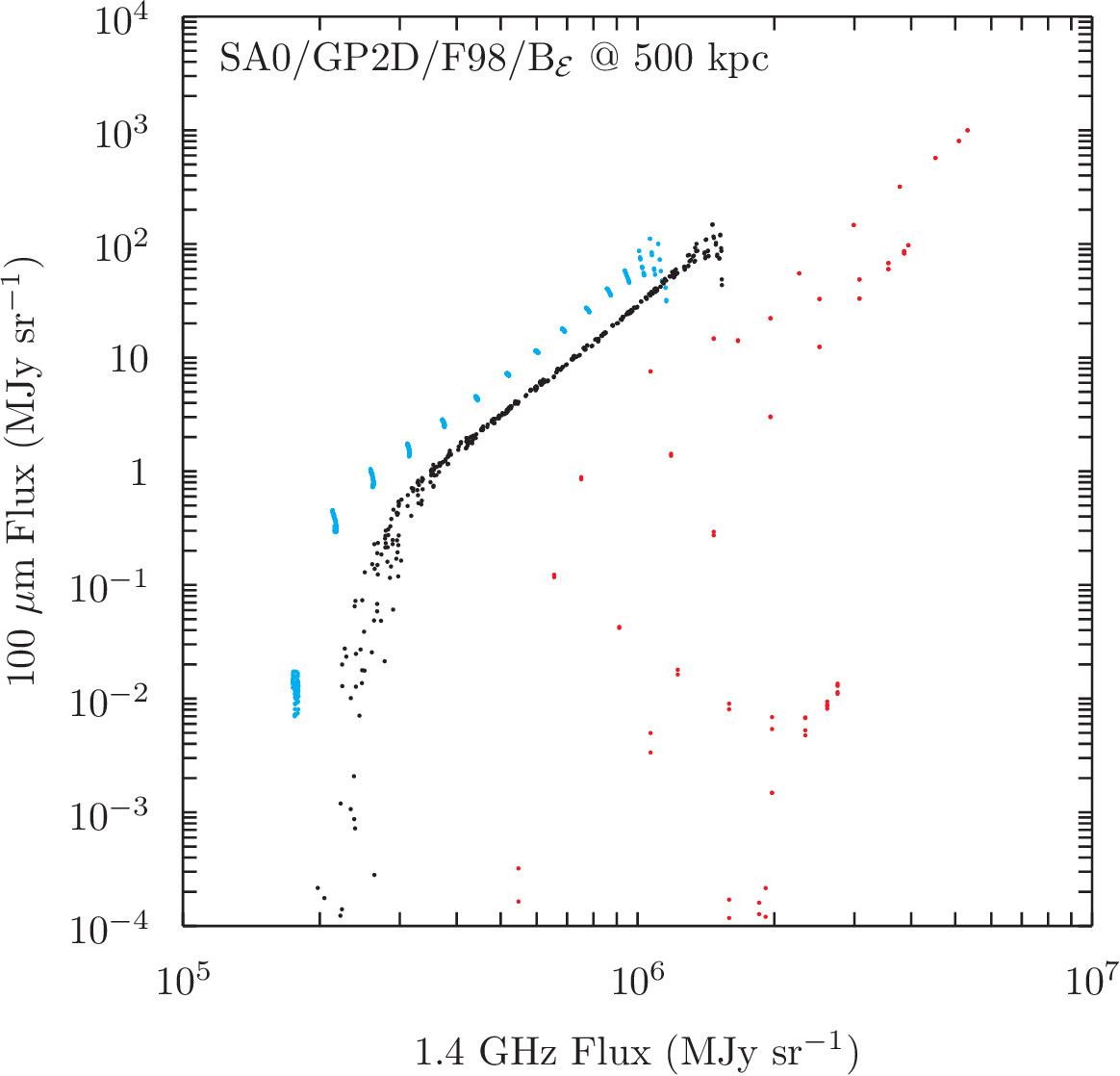}\hfill \\
  \includegraphics[height=0.47\textwidth]{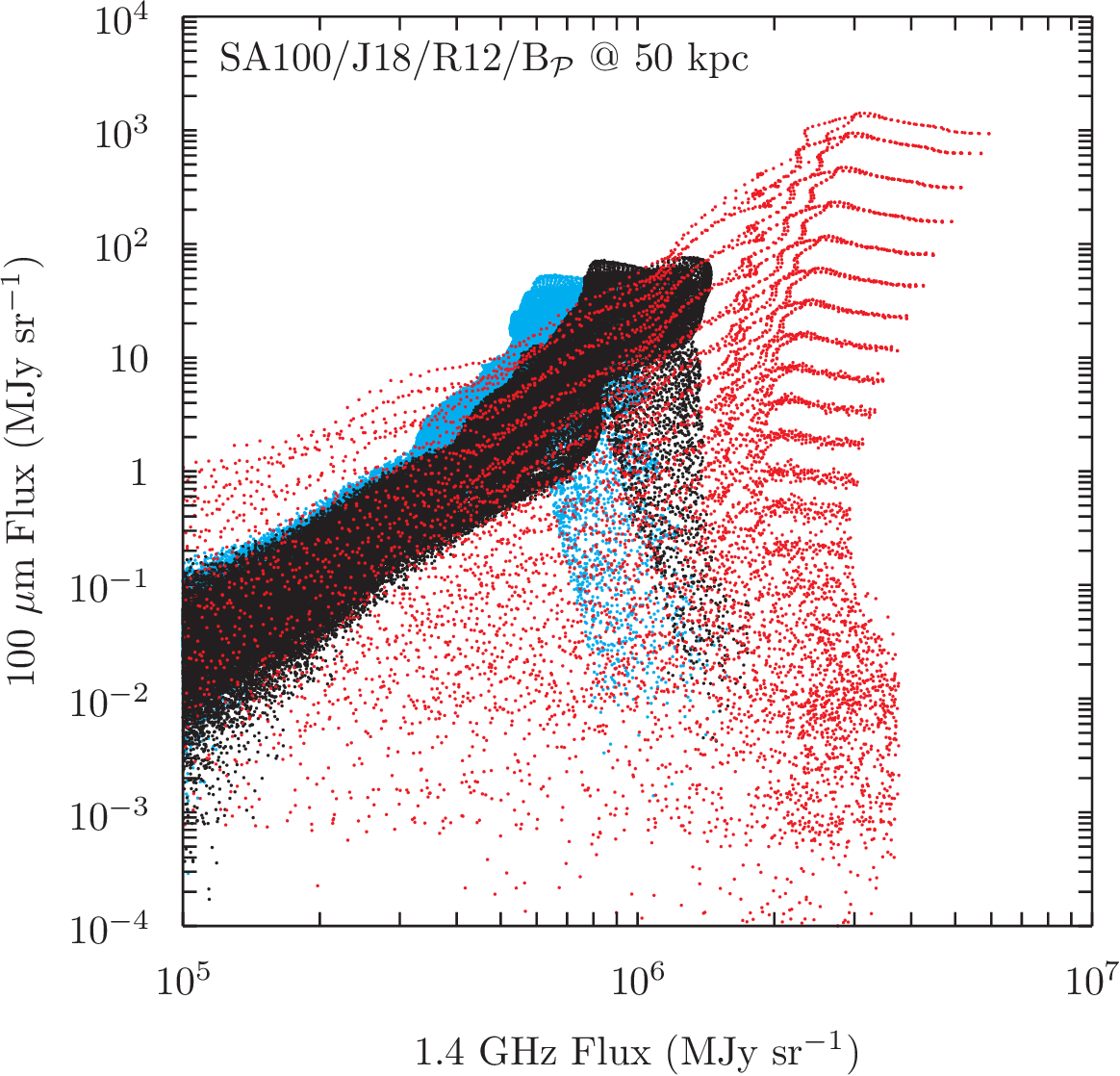}\hfill 
  \includegraphics[height=0.47\textwidth]{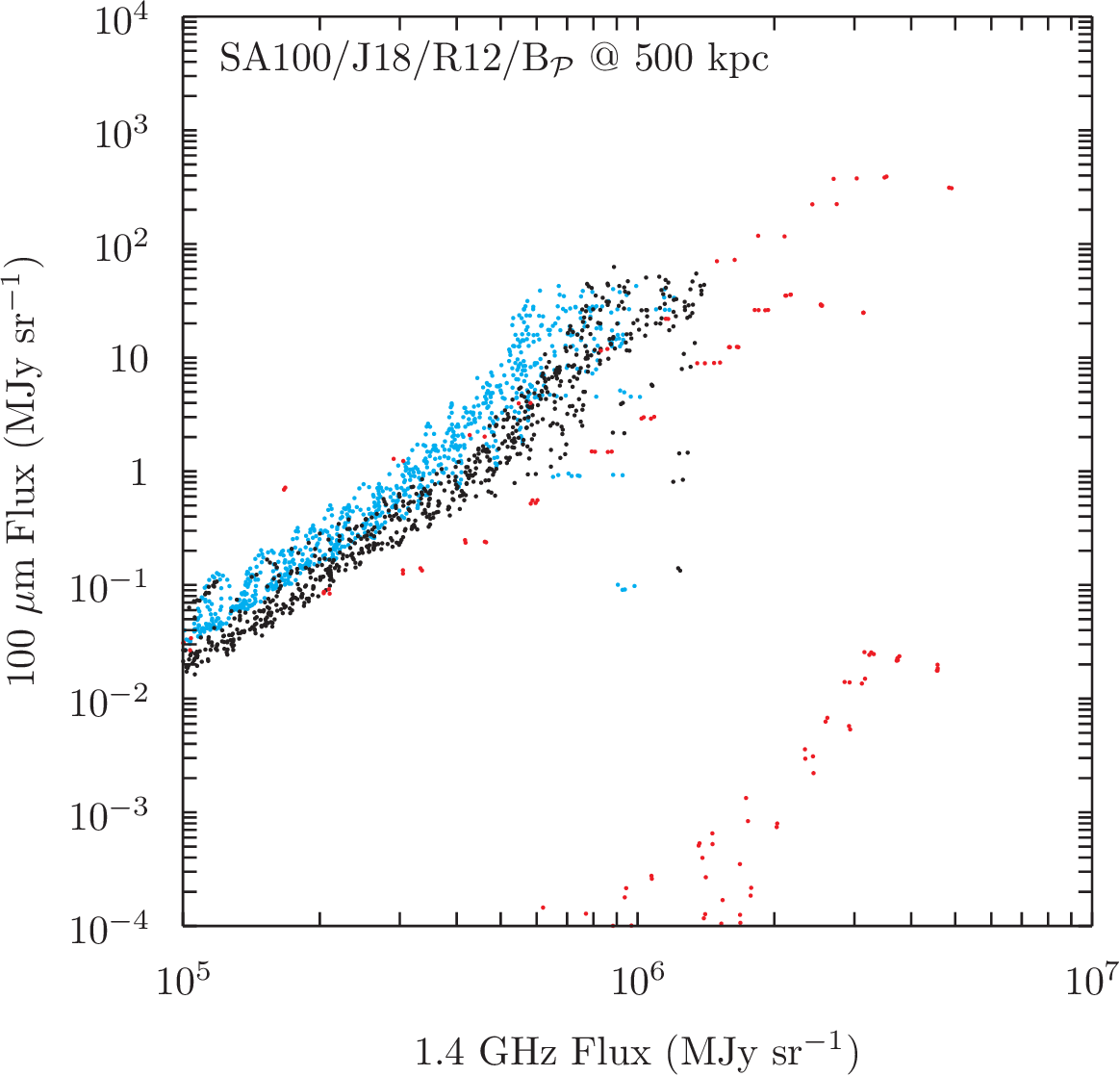}\hfill
  \caption{Correlation plots of 1.4~GHZ flux vs.\ 100$\mu$m flux for
    the SA0-GP2D-F98-B$_\mathcal{E}$ (top) and SA100-J18-R12-B$_\mathcal{P}$ (bottom) configurations for observer located at 50~kpc (left) and 500~kpc (right) from the MW centre looking face-down (cyan), at 45$^\circ$ inclination (black), and 90$^\circ$ inclination (red). \label{fig:ext_obs_irradio} }
\end{figure*}

The SA50 model combines these effects, producing its distinct plateau. Its inter-arm regions contain protons from two sources: the diffuse population that has propagated from the arms and a smooth ``floor'' from the disc-like source component. The hadronic \gray{} emission in these regions is therefore produced by both populations, resulting in a relatively high and uniform flux.
In contrast, the radio emission is primarily generated by the electron ``floor'' from the disc component, as the energetic electrons from the arms are less likely to reach these regions.
Consequently, the inter-arm regions manifest as a population of pixels where the \gray{} flux is high and nearly constant, while the radio flux declines radially with the disc profile.
It is this decoupling of the mixed-origin proton population from the single-origin electron population that produces the observed plateau for the SA50 configuration.

The influence of observational geometry is most apparent when varying the observer's inclination, as shown in Figs.~\ref{fig:ext_obs50kpc_hp6_45deg} and~\ref{fig:ext_obs50kpc_hp6_edge}.
At a 45$^\circ$ inclination (Fig.~\ref{fig:ext_obs50kpc_hp6_45deg}), the quasi-linear relationships seen in the face-on view are preserved for both the RC--\gray{} (top panels) and IR--\gray{} (bottom panels) correlations.
The LOS integration geometry still effectively averages the different galactic components, from the dense disc to the extended halo.
This maintains the fundamental scaling between the emission bands, resulting in a fairly tight correlation that is simply shifted to higher intensities due to the longer path lengths through the inclined disc.

In contrast, the simple correlation is no longer observed in an edge-on view (90$^\circ$ inclination), as demonstrated in Fig.~\ref{fig:ext_obs50kpc_hp6_edge}. For this inclination angle, the geometric projection separates rather than averages the emission components.
Sightlines confined within the galactic plane pass through very high column densities of gas and dust, strongly enhancing hadronic and IR emission.
Conversely, sightlines at higher galactic latitudes probe the extended halo of CRs, magnetic fields, and ISRF, but miss the dense disc almost entirely.
This geometric decoupling produces distinct, bifurcated tracks in the correlation plots for both the RC--\gray{} and IR--\gray{} relationships, disrupting the simple scaling seen from other perspectives. 

The breakdown is particularly pronounced for the IR--\gray{} relationship. The IR-emitting dust is the most vertically confined component in the models, appearing as a thin, bright line in an edge-on view.
Consequently, most LOS path lengths have zero IR flux, while the few that do intersect the disc have extremely high IR and hadronic \gray{} emission.
The result is a sparsely populated, scattered distribution with no clear linear trend.
Meanwhile, the RC--\gray{} correlation also shows increased scatter and bifurcation, but the effect is less extreme.
The quasi-linear relationship is still evident for the RC--IC-emission component in the near-axisymmetric case SA0-GP2D-F98-B$_\mathcal{E}$.
The radio-emitting synchrotron halo is more vertically extended than the dust disc, leading to a less severe geometric decoupling between the radio, the IC \gray{} emission component and also the total \gray{} emission.

Figures~\ref{fig:ext_obs500kpc_hp9_sa0} and~\ref{fig:ext_obs500kpc_hp9_sa100} shows correlation plots for observer located 500 kpc from the MW centre for the SA0-GP2D-F98-B$_\mathcal{E}$ and SA100-J18-R12-B$_\mathcal{P}$ configurations, respectively.
Each figure has the observer viewing the Galaxy face-on (left), at an inclination of 45$^\circ$ (middle), and edge-on (right).
This distance is somewhat smaller than that to M31, but is giving an idea for how the correlations are altered when individual pixels average over larger regions.
We use the same HEALPix resolution, $N_{\rm side}$=512, corresponding to a spatial scale of $\sim$1 kpc at this distance.

The face-on and 45$^\circ$ correlations for the SA0-GP2D-F98-B$_\mathcal{E}$ configuration at 500 kpc (Fig.~\ref{fig:ext_obs500kpc_hp9_sa0}) are similar to the 50~kpc case.
This stability is expected, because degrading the resolution over a smooth radial profile does not fundamentally alter the scaling.
The critical change is seen in the edge-on perspective.
The bifurcated structures present at 50 kpc are greatly reduced and replaced by a tighter, quasi-linear correlation.
This occurs because the $\sim$1~kpc pixel scale is now too large to resolve the vertical stratification of the disc and halo, and the resulting spatial averaging enforces a linear relationship.

Figure~\ref{fig:ext_obs500kpc_hp9_sa100} demonstrates this averaging effect using the complex, structured SA100-J18-R12-B$_\mathcal{P}$ configuration.
In the face-on and inclined views, the significant scatter caused by the arm-interarm contrast at 50 kpc is reduced.
The $\sim$1~kpc pixel scale is larger than the typical arm separation, effectively averaging over these features and removing the source of the scatter.
The edge-on view shows the combined effect of this horizontal averaging with the vertical averaging seen in Fig.~\ref{fig:ext_obs500kpc_hp9_sa0}.
The process simultaneously smooths over both the spiral structures and the distinct vertical layers of the disc and halo.
The correlation is not only restored but also tightened, making the observed relationship for this configuration very similar to the simpler nearly axisymmetric model for the same resolution.

Figure~\ref{fig:ext_obs_irradio} shows the corresponding RC--IR correlation plots at 50 and 500~kpc for the SA0-GP2D-F98-B$_\mathcal{E}$ and SA100-J18-R12-B$_\mathcal{P}$ configurations at face-on, 45$^\circ$, and 90$^\circ$ inclination angles.
At 50~kpc (left panels), the resolved structures in the correlation plots are a direct diagnostic of the underlying galactic models and viewing geometry. For the SA0-GP2D-F98-B$_\mathcal{E}$ configuration, the face-on and 45$^\circ$ views produce an extremely tight, linear relationship, as the radio and IR emissions both trace the same smooth radial profile.
The first hint of complexity appears in the edge-on view (red), where the correlation shows a strong broadening.
This is the signature of geometric decoupling, as the high-resolution view begins to resolve the differing vertical scale heights of the thin dust disc and the more extended synchrotron-emitting halo.
This effect is far more pronounced for the SA100-J18-R12-B$_\mathcal{P}$ configuration.
Here, even the face-on view exhibits significant intrinsic scatter.
This scatter is driven by the resolved arm-interarm contrast, where the main axis of the correlation is set by the high-flux arms and low-flux interarm regions, while the scatter around this axis reflects the imperfect local correspondence between the gas/dust and the smoother magnetic field.
For this model, the edge-on view shows the most pronounced scatter, as it superimposes the strong vertical decoupling of the disc and halo onto the already complex spiral structure, creating a messy, poorly-correlated point cloud.

In contrast to this resolved complexity, the 500~kpc plots (right panels of Fig.~\ref{fig:ext_obs_irradio}) demonstrate the homogenising effect of spatial averaging.
The $\sim$1~kpc pixel scale is too large to resolve any of the features seen at 50~kpc.
It simultaneously averages over both the vertical stratification of the dust disc and radio halo, which removes the edge-on scatter, and the horizontal arm-interarm contrast, which eliminates the intrinsic scatter in the structured model.
Consequently, the distinct, model- and inclination-dependent point clouds all collapse onto a single, quasi-linear trend.
At this low resolution, the correlations for the simple axisymmetric and complex structured models become very similar.

\section{Discussion}

The results of this study indicate that the observed tightness of the radio-IR-\gray{} correlations is primarily a consequence of viewing geometry, but the mechanism depends on inclination. In face-on and moderately inclined systems, a tight correlation arises because the emitting components, the gas, magnetic fields, and the ISRF weighted by the CR distributions, all share a common large-scale radial decline. Each pixel, while integrating through the relatively shallow vertical depth of the disc, effectively averages over a region where all components are at a similar point on their respective radial gradients.
This spatial averaging naturally produces a quasi-linear mapping between observables, even when the underlying CRs are not locally calorimetric.

Meanwhile, at high inclinations, the geometry acts as a stratification tool.
The long LOS path-lengths can be separated into those passing through the thin, IR-bright and gas/$\pi^0$-decay \gray{} emitting disc and those passing only through the more extended radio- and \gray{}-emitting halo.
This geometric decoupling is particularly pronounced for the IR component where the correlation is completely broken producing the wide scatter seen for our models where sightlines with high IR flux are distinct from those with little or no IR flux (e.g., bottom panels of Fig.~\ref{fig:ext_obs50kpc_hp6_edge}). That the correlation's tightness is so dependent on viewing geometry strongly suggests the observed scaling is a default outcome of viewing a radially structured disc, not a direct tracer of local physics. Therefore, the inference of calorimetry from a tight correlation alone is not uniquely determined, and the more diagnostic information is contained in the second-order features: the scatter, non-linearities, and how the correlation breaks down with inclination and resolution.

Deviations from a quasi-linearity are also evident in our models. These arise because the relative distributions of CRs and the target ISM density fields are not constant across the MW.
Different galactic environments, such as the inner and outer MW, exhibit distinct local scaling relations.
Our pixel-by-pixel plots sample all these regions, and the break in the correlation's slope marks the transition between these large-scale regimes. The sensitivity of this curvature to the assumed large-scale distributions is evident when comparing our SA50 and SA100 source models (Fig.~\ref{fig:ext_obs50kpc_hp6_srcvar}), or varying the different ISM components in Fig.~\ref{fig:ext_obs50kpc_hp9_radiogamma}.
These departures from linearity therefore trace the large-scale galactic structure, revealing information that is lost in unresolved, galaxy-integrated observations.

The inclination dependence of the correlations provides clear evidence for this geometric effect, offering a physical basis for interpreting observations of nearby resolved galaxies.
As our modelling shows, the specific diagnostic power depends on the viewing angle.
An edge-on or highly inclined view reveals vertical stratification, enabling to separate disc and halo components provided the system is sufficiently resolved.
But the best resolved system (LMC) is only at a moderate inclination $\approx$$35^\circ$, so the major information is from the analysis of the scatter and non-linearities within the correlation plane.
Because of its well defined disc-like morphology this makes the LMC an ideal laboratory to establish a genuine sub-galactic correlation, and test how local physics deviates from the global scaling behaviour.
However, this has not so far not been done.
Analyses to date employ global template fitting \citep[e.g.,][]{2010A&A...512A...7A, 2015ApJ...808...44F, 2016A&A...586A..71A}, which conflate emission from physically distinct regimes: the large-scale projection effect from the disc's radial gradient, localised CR enhancements from recent injection surrounding star-formation regions, and the quiescent diffuse ISM. A good fit to a global template does not, therefore, equate to a physical understanding.

To move beyond the limitations of global templates, a morphologically-guided segmentation of the emission using multi-wavelength data to isolate distinct physical environments could be made.
Masks could be defined for the regions of intense, recent star formation (e.g., 30 Doradus, N11), supergiant shells identified in H$\alpha$ and radio surveys, and the remaining, more quiescent diffuse ISM.
Applying the correlation analysis separately to the pixels within each environment would allow direct test for the geometric hypothesis.
The quiescent ISM should exhibit a tight, linear correlation, confirming it as the domain where the projection effect dominates.

Meanwhile, the young star-forming complexes, where the CRs are freshly injected, might show a steeper slope or less scatter, approaching the idealised calorimetric limit.
The supergiant shells and older structures could reveal signatures of CR transport, such as the hadronic plateau at low radio fluxes predicted by our SA50 model, which arises from the spatial decoupling of far-propagating protons from more rapidly cooling electrons.
The successful correlation analysis of the 30 Doradus region by \citet{2012ApJ...750..126M} serves as a proof-of-concept for this approach, which could be extended to other distinct regions to move beyond global template fitting toward a physical diagnosis of CR propagation on sub-kpc scales.

The SMC offers a contrasting, and arguably more complex, testbed for our geometric hypothesis.
Unlike the LMC, the SMC is firmly established as a non-calorimetric or ``leaky'' galaxy, with \textit{Fermi}-LAT observations implying that a significant fraction of its CRs escape before losing their energy \citep{2018ApJ...867...44L}.
This makes it an ideal case of our predictions: any observed multi-wavelength correlation in the SMC cannot be the result of local calorimetry and must therefore be driven by other factors, such as the projection effects we investigate. 

The challenge and opportunity lie in its geometry.
The traditional view of the SMC as a simple, highly-inclined disc is now being replaced by a much more intricate picture.
Recent analysis suggests the SMC is composed of two distinct, massive star-forming systems separated by $\sim$5~kpc along the LOS, each with its own chemical composition \citep{2024ApJ...962..120M}. This superimposed geometry invalidates any simple inclination-based interpretation and instead presents a scenario more akin to the stratified, multi-component systems explored in our edge-on models, but realised in a far more complex 3D arrangement.
Different sightlines will intersect one or both of these structures, leading to a superposition of emission from physically distinct environments.
Such a configuration should produce significant, structured scatter in the correlation plots, directly reflecting the complex projection. With the recent, careful separation of thermal and non-thermal radio emission in the SMC \citep{2022MNRAS.510...11H}, the data are now in place to perform this critical test. The SMC thus provides a unique laboratory to move beyond simple geometric models and test whether the scatter in the radio-IR-\gray{} correlation plane is a direct diagnostic of a known, complex underlying galactic structure.

For M31, its viewing angle of $\approx$$77^\circ$ tests the high-inclination regime where our models predict that geometric stratification should dominate over in-plane correlations. The central prediction is a decoupling between the thin, thermal/IR-emitting disc and a more vertically extended non-thermal component. This physical stratification has been directly observed: radio surveys show that the synchrotron-emitting ring in M31 is significantly wider and has a larger scale length than the thermal one \citep{2020A&A...633A...5B}. This geometric decoupling naturally explains the finding that the synchrotron-IR correlation is weak on sub-kpc scales \citep{2013A&A...557A.129T}. The smoothness of the non-thermal emission, a result of the CR diffusion away from star-forming regions \citep{2021A&A...651A..98F}, further breaks the local correspondence with clumpy IR-traced star formation. While a detailed pixel-by-pixel \gray{} correlation remains challenging, the underlying mechanism is already validated with the RC--IR. The observed weak correlations in M31 are not an anomaly, but a predicted consequence of its viewing geometry.
This is confirming that the breakdown of the correlation is as diagnostic as its tightness.

The limitation of low \gray{} photon statistics for all but the nearest galaxies can be circumvented by testing our predictions on the RC-IR correlation, where larger samples exist.
In particular, the recent analysis of CHANG-ES galaxies \citep{2012AJ....144...43I,2024Galax..12...22I} by \citet{2025A&A...699A.243H} empirically decomposes the vertical radio emissions of 22 edge-on systems into thin ($\sim$0.45~kpc) and thick ($\sim$1.5~kpc) synchrotron discs. After deprojecting to find the intrinsic properties, they show that the thin synchrotron disc's intensity correlates strongly with the SFR surface density (the IR proxy), while the thick synchrotron disc shows a much weaker, scattered correlation.

This finding provides the direct observational basis for the breakdown seen in our edge-on models. The observed scatter in a highly inclined galaxy is not random but is the result of superimposing these two physically distinct populations along the LOS: sightlines through the thin, IR-bright disc follow a tight correlation, while those probing only the extended, weakly-correlated radio halo do not. The \citet{2025A&A...699A.243H} study, by successfully isolating and measuring the different behaviours of these components, provides the empirical demonstration that the breakdown of a simple correlation is a direct and predictable consequence of vertical galactic stratification.

Our results are derived from simulations of a single MW-like system. They provide a strong framework for interpreting correlations within spiral galaxies but cannot be directly extrapolated to the global radio-IR-\gray{} correlation. Our models do not cover the diverse range of galactic systems, and understanding the role of geometry across all galaxy types remains a topic for future investigation.

  We propose, however, a new interpretation for the positions of unresolved normal star-forming galaxies on the radio-IR-\gray{} correlation. If geometric averaging is a dominant effect, the diagnostic power lies not in the correlation's tightness, but in the scatter around it. In this view, the scatter is not random noise but a direct physical signal. Clear outliers would signify distinct physical galactic states: either ``leaky'' systems with a porous ISM or compact starbursts with enhanced CR trapping.
This approach turns a simple scaling relation into a tool for finding second-parameter dependencies that link a galaxy's outlier status to its physical properties.

\section{Summary}

Using 3D simulations of the MW, this work argues that the tight radio-IR-\gray{} correlation in star-forming galaxies is primarily an emergent property of geometric projection, not a direct signature of CR calorimetry.
We find that the correlation's behaviour is mainly dictated by viewing angle. In face-on and moderately inclined systems, LOS integration through a radially-structured disc naturally produces a linear correlation, even when local calorimetry is absent.
Conversely, for highly-inclined, edge-on systems, the correlation becomes scattered and weak. This is because sightlines passing through the plane are distinct from those probing the halo, separating the thin, IR-emitting disc from the more extended radio and \gray{} emission.

This geometric picture reframes the interpretation of observations. For resolved, moderately-inclined galaxies like the LMC, our study suggests that the scatter around the main correlation is the primary diagnostic tool, flagging regions where local physics deviates from the global projection. For highly-inclined systems like M31, the breakdown of the correlation itself becomes the key diagnostic, confirming the geometric decoupling of emission components.

Finally, we have shown that at low resolution, where internal structures are blurred, our models always recover a tight correlation regardless of inclination. This result reinforces a key finding of our work: while spatial averaging can enforce a tight correlation, the physically diagnostic information is contained within the scatter and structural features that are only visible at high resolution, which directly trace the interplay between galactic structure and CR propagation in the modelled system.

\acknowledgements
TAP and IVM acknowledge partial support from NASA Grants No.~80NSSC23K0169,
80NSSC22K0718, and 80NSSC22K0477.

\bibliography{radioirgamma,lowecr,gp_v57,imos,imos_icrc2015,ms_iso}

\end{document}